\documentclass[twocolumn,twocolappendix,tighten,dvipsnames]{aastex63} 

\usepackage{eht,amsmath}
\usepackage[nolist,nohyperlinks]{acronym}
\usepackage{tabularx}
\newcolumntype{C}{>{\centering\arraybackslash}X}
\graphicspath{{./}{figures/}}
\usepackage{comment}

\usepackage{savesym}
\savesymbol{tablenum}
\usepackage{siunitx}

\newcommand\dha[1]{{\color{violet}[dh: #1]}}

\newcommand\hb[1]{{\color{BurntOrange}[hb: #1]}}

\newcommand\old[1]{{\color{lightgray}[old: #1]}}

\newcommand{\sdiam}{$\sim$50\,$\mu$as\xspace}
\begin{document}

\begin{acronym}
\acro{sn}[$S/N$]{signal-to-noise ratio}
\acro{vlbi}[VLBI]{very-long-baseline interferometry}
\acroplural{vlbi}[VLBI]{Very long baseline interferometry}
\acro{grmhd}[GRMHD]{general relativistic magnetohydrodynamics}
\acro{mhd}[MHD]{magnetohydrodynamics}
\acro{as}[as]{arcseconds}
\acro{agn}[AGN]{Active Galactic Nuclei}
\acro{llagn}[LLAGN]{low-luminosity AGN}
\acro{cena}[Cen~A]{Centaurus A}
\acro{sgra}[\sgra]{Sagittarius~A*}
\acro{agn}[AGN]{active galactic nuclei}
\acro{eht}[EHT]{Event Horizon Telescope}
\acro{bhc}[BHC]{\href{https://blackholecam.org}{BlackHoleCam}}
\acro{tanami}[TANAMI]{Tracking Active Galactic Nuclei with Austral Milliarcsecond Interferometry}
\acro{smbh}[SMBH]{supermassive black hole}
\acroplural{smbh}[SMBHs]{supermassive black holes}
\acro{aa}[ALMA]{Atacama Large Millimeter/submillimeter Array}
\acro{ap}[APEX]{Atacama Pathfinder Experiment}
\acro{pv}[PV]{IRAM~30\,m Telescope on Pico Veleta}
\acro{jc}[JCMT]{James Clerk Maxwell Telescope}
\acro{lm}[LMT]{Large Millimeter Telescope Alfonso Serrano}
\acro{sp}[SPT]{South Pole Telescope}
\acro{sm}[SMA]{Submillimeter Array}
\acro{az}[SMT]{Submillimeter Telescope}
\acro{chandra}[Chandra]{Chandra X-ray Observatory}
\acro{c}[c]{speed of light}
\acro{pc}[pc]{parsec}
\acro{gr}[GR]{general relativity}
\acro{aips}[\textsc{aips}]{\href{http://www.aips..edu}{Astronomical Image Processing System}}
\acro{casa}[CASA]{\href{https://casa..edu}{Common Astronomy Software Applications}}
\acro{symba}[\textsc{symba}]{\href{https://bitbucket.org/M_Janssen/symba}{SYnthetic Measurement creator for long Baseline Arrays}}
\acro{rpicard}[\textsc{Rpicard}]{\href{https://bitbucket.org/M_Janssen/picard}{Radboud PIpeline for the Calibration of high Angular Resolution Data}}
\acro{hops}[HOPS]{\href{https://www.haystack.mit.edu/tech/vlbi/hops.html}{Haystack Observatory Postprocessing System}}
\acro{hbt}[HBT]{Hanbury Brown and Twiss}
\acro{tov}[TOV]{Tolman-Oppenheimer-Volkoff}
\acro{mri}[MRI]{magnetorotational instability}
\acro{adaf}[ADAF]{advection-dominated accretion flow}
\acro{adios}[ADIOS]{adiabatic inflow-outflow solution}
\acro{cdaf}[CDAF]{convection-dominated accretion flow}
\acro{bz}[BZ]{Blandford-Znajek}
\acro{bp}[BP]{Blandford-Payne}
\acro{em}[EM]{electromagnetic}
\acro{blr}[BLR]{broad-line region}
\acro{nlr}[NLR]{narrow-line region}
\acro{ism}[ISM]{interstellar medium}
\acro{edf}[eDF]{electron distribution function}
\acro{pic}[PIC]{particle-in-cell}
\acro{sane}[SANE]{standard and normal evolution}
\acro{mad}[MAD]{magnetically arrested disk}
\acro{jive}[JIVE]{\href{http://www.jive.eu}{Joint Institute for VLBI ERIC}}
\acro{}[NRAO]{\href{https://www.nrao.edu}{National Radio Astronomy Observatory}}
\acro{muas}[\textmu as]{microarcseconds}
\acroplural{agn}[AGN]{Active galactic nuclei}
\acro{jy}[Jy]{jansky}
\acro{pa}[PA]{position angle}
\acro{srmhd}[SRMHD]{special relativistic magnetohydrodynamics}
\acro{fov}[FOV]{field of view}
\acro{sefd}[SEFD]{system equivalent flux density}
\acroplural{sefd}[SEFDs]{system equivalent flux densities}
\acro{ALMA}[ALMA]{Atacama Large Millimeter/submillimeter Array}
\acro{SMA}[SMA]{Submillimeter Array}
\acro{CARMA}[CARMA]{Combined Array for Research in Millimeter-wave Astronomy}
\acro{snr}[$S/N$]{signal-to-noise ratio}
\end{acronym}

\def\swift{Swift\xspace}
\def\chandra{Chandra\xspace}
\def\nustar{NuSTAR\xspace}

\nocite{PaperI}
\nocite{PaperII}
\nocite{PaperIII}
\nocite{PaperIV}
\nocite{PaperV}
\nocite{PaperVI}

%







\title{First Sagittarius A* Event Horizon Telescope Results. II.\\ 
EHT and Multi-wavelength Observations, Data Processing, and Calibration}
\author{The Event Horizon Telescope Collaboration}

\shorttitle{EHT \sgra Paper II}
\shortauthors{EHTC et al.}


\begin{abstract} 
We present Event Horizon Telescope (EHT) 1.3\,millimeter measurements of the radio source located at the position of the supermassive black hole Sagittarius A* (\sgra), 
collected during the 2017 April 5--11 campaign.
The observations were carried out with eight facilities at six locations across the globe. 
Novel calibration methods are employed to account for Sgr A*'s flux variability. The majority of the 1.3\,millimeter emission arises from horizon scales, where intrinsic structural source variability is detected on timescales of minutes to hours. The effects of interstellar scattering on the image and its variability are found to be subdominant to intrinsic source structure.
The calibrated visibility amplitudes, particularly the locations of the visibility minima, are broadly consistent with a blurred ring 
with a diameter of \sdiam,  
as determined in later works in this series.
Contemporaneous multi-wavelength monitoring of \sgra was performed at 22, 43, and 86 GHz and at near infrared and X-ray wavelengths.  
Several X-ray flares from \sgra are detected by \chandra, one at low significance jointly with \swift on 2017 April 7 and the other at higher significance jointly with \nustar on 2017 April 11. The brighter April 11 flare is not observed simultaneously by the EHT but is followed by a significant increase in millimeter flux variability immediately after the X-ray outburst, indicating a likely connection in the emission physics near the event horizon. We compare \sgra's broadband flux 
during the EHT campaign to its historical spectral energy distribution and find both the quiescent and flare emission are consistent with its long-term behaviour. 
\end{abstract}


\section{Introduction}
\label{sec:intro}

The first legacy set of papers by the \ac{eht} focused on the supermassive black hole \m87 \citep[][hereafter \m87 Papers I, II, III, IV, V, and VI]{M87PaperI,M87PaperII,M87PaperIII,M87PaperIV,M87PaperV,M87PaperVI}, but 
\ac{sgra} was the spark that motivated 
the formation of the \ac{eht}.  
This work 
presents EHT observations of \sgra at 1.3\,mm, 
which serve as the foundation for the observational and theoretical papers presented in this second legacy series \citep[][hereafter Papers I, II, III, IV, V and VI]{PaperI,PaperII,PaperIII,PaperIV,PaperV,PaperVI}. 

\ac{sgra} is the black hole at the center of our own Milky Way, and is  
the only supermassive black hole observable 
at a distance of a few kiloparsecs.  
Among all known black holes, \sgra has the largest predicted angular size 
(\sdiam, see \citetalias{PaperIII} and \citetalias{PaperIV}, and references therein). It is also a highly variable source, with flickering, flares, and other stochastic processes occurring across the electromagnetic spectrum on short and long timescales. These unique characteristics make \ac{sgra} an important laboratory for studying the fundamental physics and astrophysics of black holes at high angular resolution.


\sgra has been observed with millimeter \ac{vlbi} for over a quarter century. After initial successful 1\,mm \ac{vlbi} tests on quasars \citep{1990ApJ...360L..11P,1995A&A...299L..33G}, \ac{sgra} was first successfully detected on a \ac{vlbi} baseline between the IRAM 30-m telescope on Pico Veleta (PV) in Spain and a single antenna of the Plateau de Bure Interferometer in France in 1995 \citep{1997A&A...323L..17K}. This detection revealed a compact source, with a size of $\left( 110\pm60 \right)$\,\uas \citep{1998A&A...335L.106K}. Early size measurements at 3\,mm and 1\,mm were larger than expected 
\citep[cf.\  e.g.,][]{1998ApJ...508L..61L}, indicating that short-wavelength \ac{vlbi} measures the intrinsic structure in \ac{sgra} rather than interstellar scattering along the line of sight.
Subsequent VLBI experiments using wider recorded bandwidth and three telescopes with longer baselines provided a tighter estimate of the source size, $43^{+14}_{-8}$\,\uas, giving the first unambiguous detection of horizon-scale structure in \sgra \citep{Doeleman08}. Meanwhile, continued \ac{vlbi} observations at $\lambda \gsim 3\,{\rm mm}$ were better able to characterize the properties of the anisotropic interstellar scattering screen \citep[e.g.,][]{2004Sci...304..704B,2018ApJ...865..104J}. 

Excitement from these VLBI measurements was further galvanized by crucial theoretical and technical advancements made in parallel. 
Simulations of \sgra by \citet{2000ApJ...528L..13F} demonstrated that a shadow of the sort originally predicted by \citet{1973blho.conf..215B} would be observable with millimeter-wavelength \ac{vlbi}.\footnote{More details about the appearance of black holes are given in \citetalias{M87PaperI}.} 
Technological advances greatly increased the capabilities of the growing \ac{eht}, as detailed in \citetalias{M87PaperII}.  
These advances
led to a new era in which the detection of \ac{sgra} on long baselines at 1\,mm became routine 
\citep{Fish_2011,Johnson_2015,2016ApJ...820...90F,Lu_2018}. Most significantly, the phased \ac{aa} \citep{2018PASP..130a5002M} participated in its first \ac{eht} science observations in 2017, along with other antennas that added to the baseline coverage.
Indeed, data from these observations produced the \m87 total-intensity \citepalias{M87PaperI,M87PaperII,M87PaperIII,M87PaperIV,M87PaperV,M87PaperVI} and polarization results \citep[][Papers VII and VIII hereafter]{M87PaperVII,M87PaperVIII}, as well as high angular resolution images of extragalactic radio jets \citep{Kim_2020,2021Janssen}. These data also motivate the \ac{sgra} results in this series.

As these \ac{vlbi} discoveries were advancing, \ac{sgra} was also being studied intensively at other wavelengths. Radio, millimeter, infrared, and X-ray observations showed that  
\ac{sgra} has both a very low bolometric-to-Eddington luminosity ratio of $L/L_{\rm{Edd}} \sim 10^{-9}$ \citep{Genzel10}, and a very low mass accretion rate of $\sim 10^{-9} ~\rm{to} ~10^{-7}$ M$_{\odot}$ yr$^{-1}$ \citep{Baganoff03, Marrone06, Marrone07, Shcherbakov12, YusefZadeh15}. At most wavelengths, \ac{sgra}’s flux can be decomposed into a quiescent and variable component.

In the X-ray, \sgra is a persistent source, with a flux of about $3\times10^{33}$ erg s$^{-1}$ \citep{Baganoff01, Baganoff03} from thermal bremsstrahlung radiation originating from hot plasma near the Bondi radius \citep[e.g.,][]{Quataert02, Baganoff03, Yuan03, Liu04, Wang2013}. Bright X-ray flares punctuate this emission about once per day and are characterized by non-thermal emission centered on the black hole \citep[e.g.,][]{Neilsen2013b}. 
Near infrared (NIR) detections of \sgra also reveal a highly variable source, with emission peaks observed more frequently than in the X-ray \citep{Genzel03,Ghez04,GRAVITY2020}. 
Both the X-ray and NIR variability occur on timescales of several hours, consistent with emission originating near the
black hole's innermost stable circular orbit (ISCO), 
which depends on the black hole's mass and spin.
\sgra's mid-IR flux is only marginally detected \citep[e.g.,][]{Iwata2020} or it can be inferred indirectly from model fitting. 

Millimeter polarimetry of \sgra reveals linearly polarized flux from an emitting region of $\sim10$ Schwarzschild radii ($R_S$), which indicates a dense magnetized accretion flow again extending  
out to the Bondi radius. \citet{Bower2018} find a mean rotation measure (RM) of $\sim -5\times10^5\,{\rm rad} \,{\rm m}^{-2}$ that can be modeled as a radiatively inefficient accretion flow (RIAF) with an accretion rate of $\sim 10^{-8}\,\msun\,{\rm yr}^{-1}$. 
Circular polarization is also detected at a mean value of $-1.1\pm0.2$\%. Both the RM and the circular polarization are variable on timescales of hours to months \citep{Bower2018}.

Similarly, observations of \sgra between 15 and 43 GHz reveal variability at the 5--10\% level on timescales shorter than four days \citep{Macquart06}.
\sgra's flux density distribution at 217.5, 219.5, and 234.0\,GHz 
was investigated recently by \citet{Iwata2020}; they find variability on timescales of $\sim$tens of minutes to hours, indicating that the emission at these wavelengths is also likely to arise near the ISCO.

Hence, in addition to the excitement around resolving  \ac{sgra}'s intrinsic structure at 1.3\,mm, it became clear that multi-wavelength observations during the EHT campaign would offer the first opportunity to definitively connect the black hole's variable flux components with changes observed at horizon scales. 

In this work we present the first EHT 1.3\,mm observations of \sgra, alongside multi-wavelength data collected 
contemporaneously in April 2017. 
Contemporaneous interferometric array data from ALMA and SMA  
have been analyzed and are described here \citep[and presented in more detail in a companion paper,][]{Wielgus2022}. The campaign also includes observations from 
the East Asian VLBI Network (EAVN), the Global 3\,mm VLBI Array (GMVA), the Very Large Telescope (VLT), the Neil Gehrels Swift Observatory, the Chandra X-ray Observatory, and the Nuclear Spectroscopic Telescope (\nustar). 
These coordinated observations provide (quasi-)simultaneous multi-wavelength coverage with exceptional spatial and spectral resolution. Since variability at timescales of minutes to hours can be probed 
on horizon scales by the EHT, and on a range of other spatial (and spectral) scales by these other observatories, combining them into a single ``snapshot'' spectral energy distribution (SED) maximizes the broadband constraints that the observations can place on theoretical models. 

This manuscript (\citetalias{PaperII}) is organized as follows. In Section~\ref{sec:eht_obs} we present an overview of the 2017 EHT observing campaign. Section~\ref{sec:calib} delves more deeply into the EHT data calibration and reduction specific to these \sgra data. 
Section \ref{sec:mwl_obs} outlines the multi-wavelength (MWL) campaigns that accompanied the EHT observations.
Section~\ref{sec:data} describes the resulting EHT and MWL data products, including those provided via a public data archive, and 
discusses these new observations in the context of longer-term monitoring campaigns that have characterized \sgra's variability over more than 20 years. We offer a brief summary and conclusions in Section~\ref{sec:concl}.


\section{Event Horizon Telescope Observing Campaign} 
\label{sec:eht_obs}


\begin{figure*}
\centering
\includegraphics[height=4.9cm]{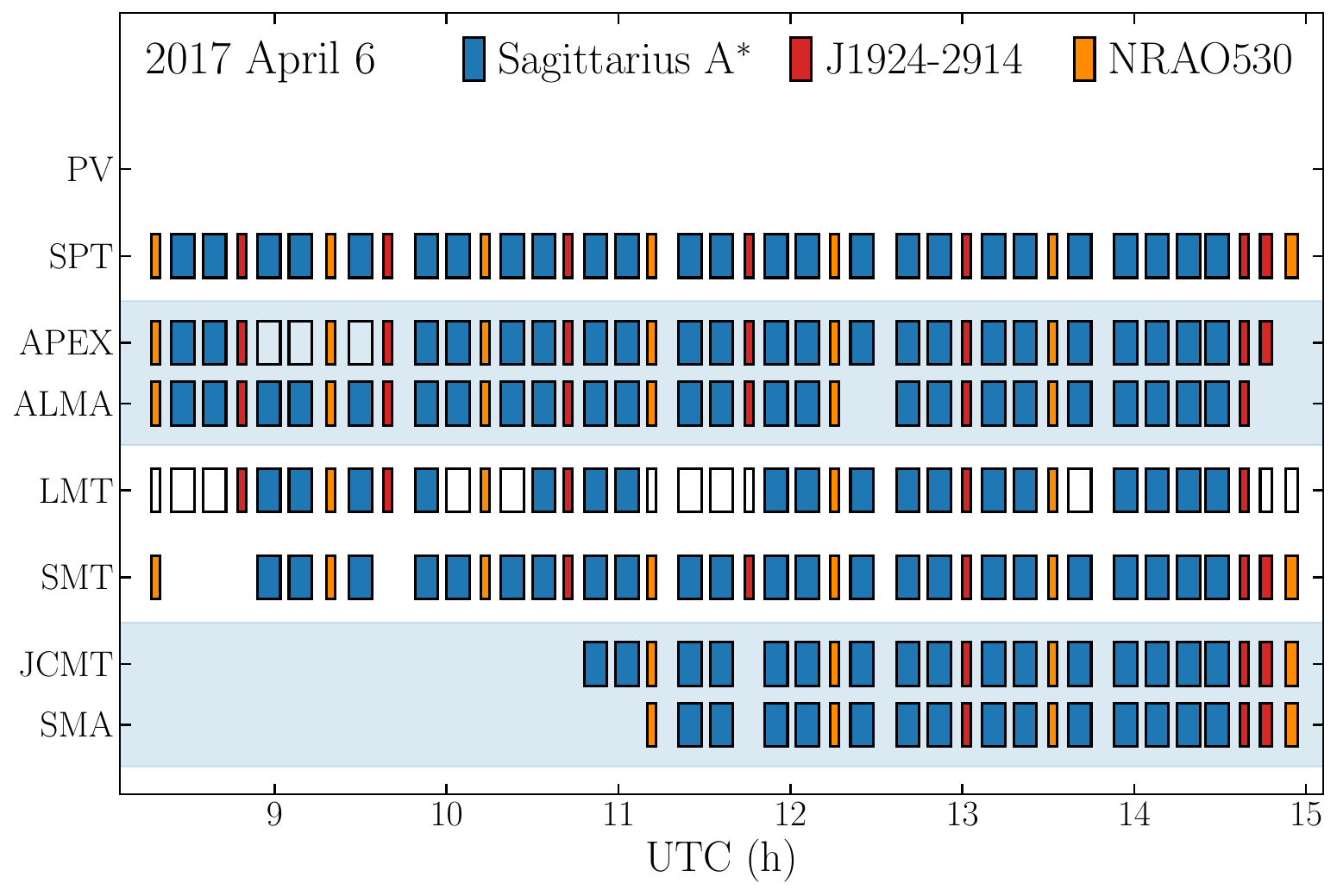}
\includegraphics[height=4.9cm]{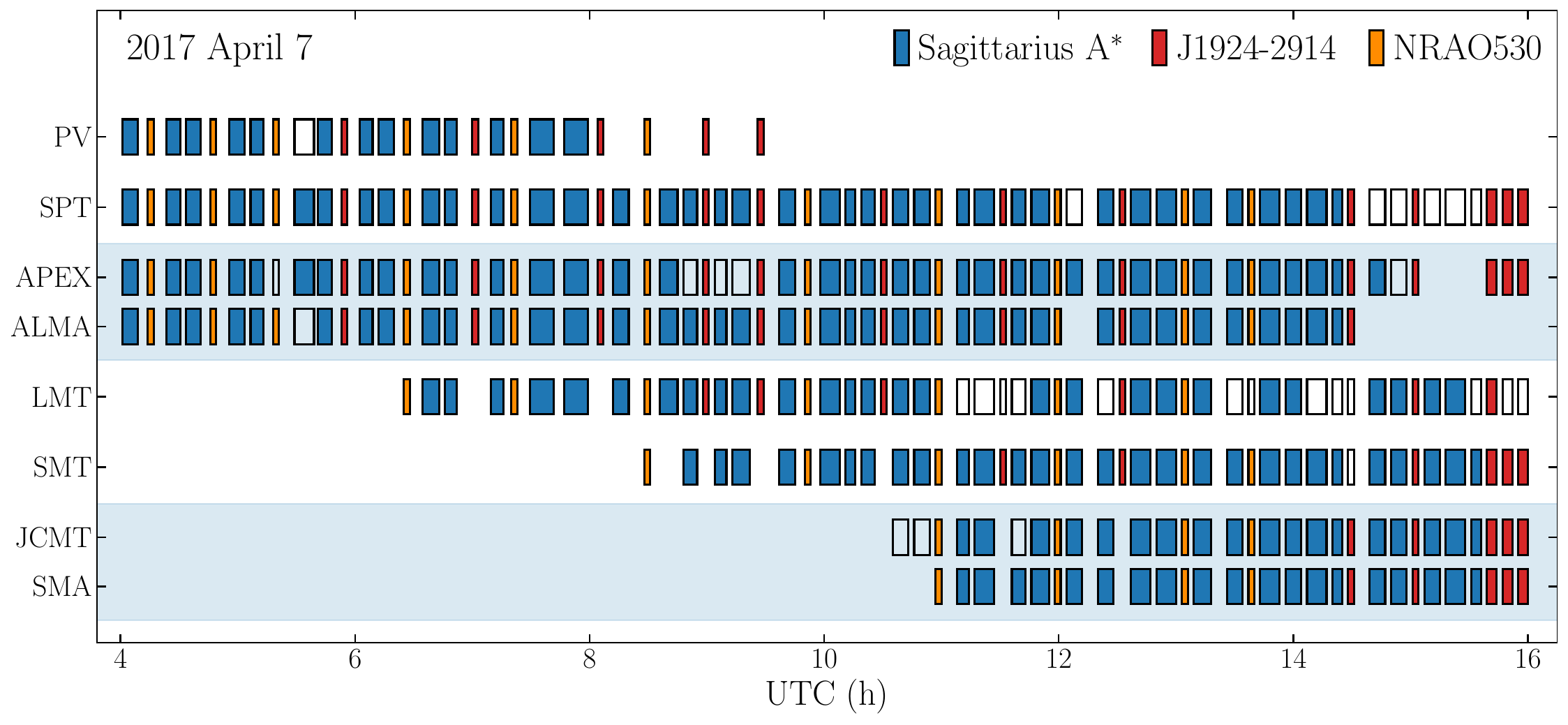}
\caption{
EHT 2017 observing schedules for \sgra and its calibrators (J1924$-$2914 and NRAO530), covering observations on April 6--7 2017. Empty rectangles represent scans that were scheduled but not observed successfully due to weather
or technical issues. The filled rectangles represent scans corresponding to detections available in the final data set. Scan durations vary between 3 and 12 minutes, as reflected by the width of each rectangle. ALMA/APEX and JCMT/SMA are pairs of co-located stations (enclosed in light blue shaded regions), providing the same $(u,v)$-coverage.
}
\label{fig:schedules}
\end{figure*}

\begin{figure*}
\includegraphics[width=\textwidth]{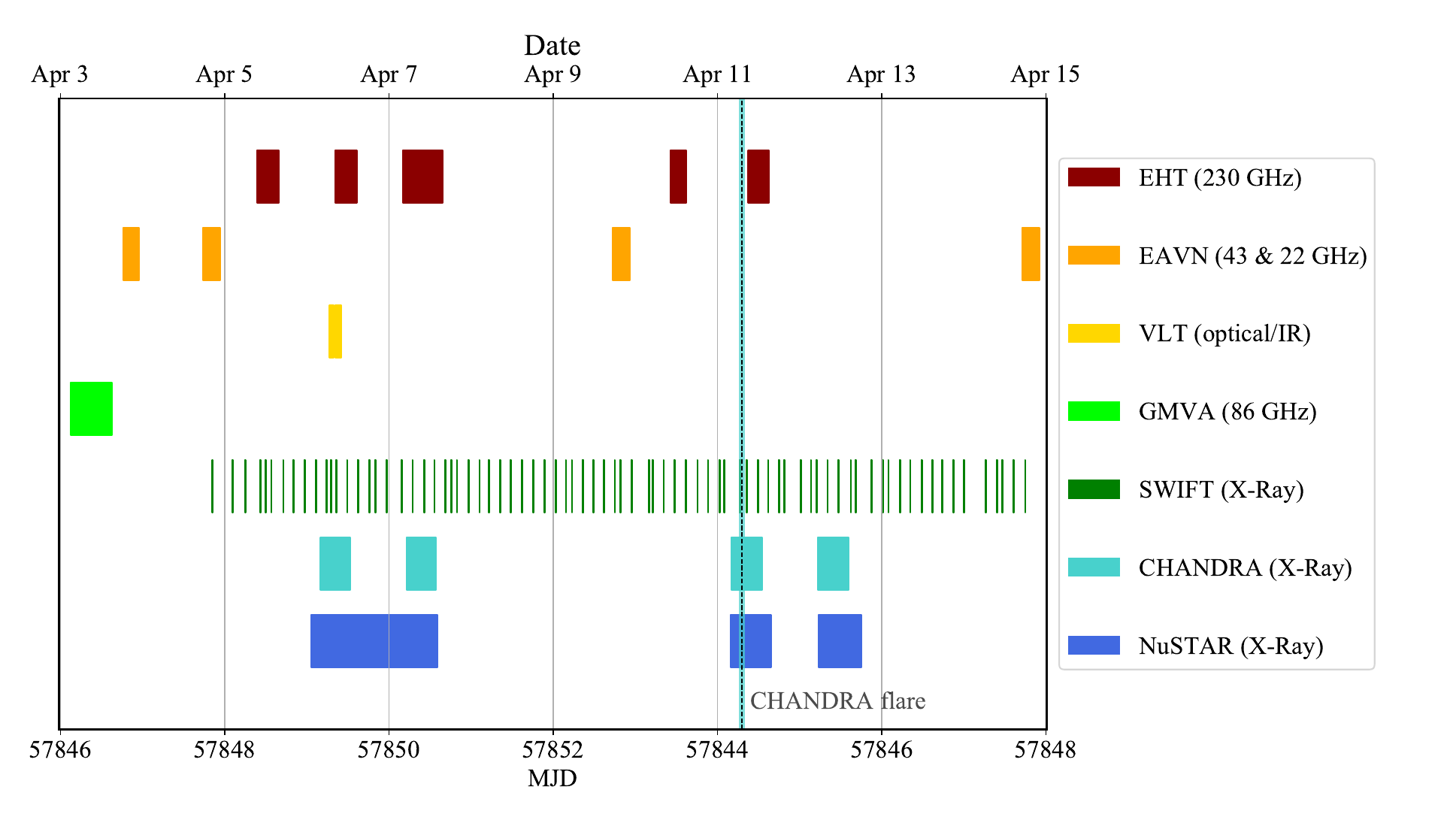}
\caption{
Instrument coverage of \sgra during the 2017 April 3--15 multi-wavelength observing campaign, which includes the East Asian VLBI Network (EAVN), the Global 3\,mm VLBI Array (GMVA), the Very Large Telescope (VLT), the Neil Gehrels Swift Observatory (SWIFT), the Chandra X-ray Observatory (CHANDRA), and the Nuclear Spectroscopic Telescope (\nustar). For the EHT data described in this work we focus on April 6 and 7, which have the best $(u, v)$-coverage and for which detailed instrument and source coverage is shown in \autoref{fig:schedules}. Details of the multi-wavelength campaign and spectral energy distribution are contained in Sections \ref{sec:mwl_obs} and \ref{subs:mwl_data}. 
\label{fig:mwl-coverage}}
\end{figure*}

The first EHT observations of \sgra were collected in April 2017, alongside contemporaneous broadband data --- the coverage is shown in Figures \ref{fig:schedules} and \ref{fig:mwl-coverage}. 
A detailed description of the EHT array and its instrumentation can be found in 
\citetalias{M87PaperII}, with further details related to the 2017 observing campaign in \citetalias{M87PaperIII}. Here we provide a brief summary of this material, along with details  
pertinent to the observations of \sgra and associated calibration sources. 

EHT observations were carried out with eight observatories at six locations: \ac{aa} and the \ac{ap} on the Llano de Chajnantor in Chile, the \ac{lm} on Volc\'{a}n Sierra Negra in Mexico, the \ac{jc} and \ac{sm} on Maunakea in Hawai`i, \ac{pv} on Pico Veleta in Spain, the \ac{az} on Mt.\ Graham in Arizona, and the \ac{sp} in Antarctica.  The locations of these telescopes are plotted in Figure~1 of \citetalias{PaperI}.

\sgra was observed on five nights: 2017 April 5, 6, 7, 10, and 11. 
ALMA did not participate in the array for observations of \sgra on 2017 April 5 or 10. \ac{pv} observed \sgra\ only on 2017 April 7.  Weather conditions were good or excellent at all sites on all five observing nights. Median opacities on each night are provided in \citetalias{M87PaperIII}.
In this series of papers, we focus our analysis on April 6 and 7, which have the best $(u, v)$-coverage.

Two bands of approximately 2\,GHz each were observed, centered at sky frequencies of 227.1 and 229.1\,GHz (``low" and ``high" bands, respectively).  Single-dish stations 
recorded a 2-bit Nyquist-sampled bandwidth of 2048\,MHz per polarization 
using Reconfigurable Open Architecture Computing Hardware 2 (ROACH2) digital backends \citep{2015PASP..127.1226V}.  The SMA observed using six or seven telescopes (depending on the observing night) whose signals were summed using the SMA Wideband Astronomical ROACH2 Machine  \citep[SWARM;][]{2016JAI.....541006P,2016past.confE...1Y}.  Since SWARM produces data in the frequency domain at a different sampling rate than at other observatories, SMA data require a post-observation, pre-correlation pass through the Adaptive Phased-array and Heterogeneous Interpolating Downsampler for SWARM (APHIDS; see also the Appendix of \citetalias{M87PaperII}).  ALMA observed as a phased array of 35--37 telescopes using the ALMA Phasing System \citep{2018PASP..130a5002M}.
Data were recorded onto Mark 6 VLBI Data Systems \citep{2013PASP..125..196W}.

All but two observatories recorded both left and right circular polarization (LCP and RCP respectively).  The JCMT was equipped with a single-polarization receiver that observed RCP on 2017 April 5--7 and LCP on 2017 April 10--11.  ALMA recorded both linear polarizations (X and Y).

The schedule consisted of scans of \sgra interleaved with scans of NRAO\,530 (J1733$-$1304) and J1924$-$2914 as calibration sources. On 2017 April 6, observations commenced after the preceding target source, M87, set at ALMA.  Generally, two eight-minute scans of \sgra were followed by a short (typically, three-minute) scan on a calibrator, with an eight-minute gap approximately every hour.  The 2017 April 7 schedule, which did not include M87, started when \sgra rose above the 20\degr\ elevation limit at ALMA.  Scan lengths on \sgra\ for this schedule were dithered between five and twelve minutes to reduce the effects of periodic sampling on detection of time variability associated with the ISCO period.  On both nights, \sgra was observed until the source set below the local horizon at the SMT and LMT, which happens at approximately the same time.  The scan coverage for these nights is shown in \autoref{fig:schedules}.

Data were correlated with the Distributed FX (DiFX) software correlation package \citep{2011PASP..123..275D} at the two correlator centers at the Max-Planck-Institut f\"{u}r Radioastronomie in Bonn, Germany and MIT Haystack Observatory in Westford, Massachusetts; for details see \citetalias{M87PaperIII}.
The CALC model was used for an a priori correction of rates and delays.
Multiple correlation passes were required to diagnose and mitigate data issues, as discussed in the Appendix of \citetalias{M87PaperIII}; parameters hereafter refer to those for the final correlation (Rev7) used for science-release data.  The final correlation produced 32 baseband channels, each 58\,MHz wide with a spectral resolution of 0.5\,MHz, and averaged to a 0.4\,s accumulation period.  The \sgra
correlation center was set to $\alpha_\mathrm{J2000} = 17^\mathrm{h}45^\mathrm{m}40\fs0356, \delta_\mathrm{J2000} =  -29\degr00\arcmin28\farcs240$\footnote{This position for \sgra is also adopted in the multi-wavelength analysis that follows.}, based on the position of \citet{Reid2004} corrected to the epoch of observation for the apparent motion introduced by the orbit of the solar system around the Galaxy. The corrected position produced smaller residual delays and rates compared with the uncorrected position, resulting in a minor improvement in sensitivity.  Subsequent to correlation, PolConvert \citep{martividal2016} was run to convert the mixed-polarization data products (XL, XR, YL, YR) to the circular basis on ALMA baselines, as described in \citet{2019PASP..131g5003G}.


\section{EHT Data Calibration and Reduction}
\label{sec:calib}

In this section, we summarize the EHT data calibration pathway and highlight aspects that are particular for the \ac{sgra} data.
A comprehensive description of the EHT data reduction methods, combined with a recap of \ac{vlbi} data calibration fundamentals, is given in \citetalias{M87PaperIII}.

\subsection{Processing pipelines}
\label{sec:calib_pipelines}

To reduce the volume of the data and to accumulate \ac{sn}, we average the visibilities in time and frequency.
To avoid non-closing errors, we must stabilize the signal by removing all significant external data corruption effects that have not been captured by the correlator model beforehand.
We apply scaling corrections as a function of time to the visibilities to ensure unity auto-correlations.
Additionally, phase errors
induced by atmospheric turbulence on ~second time scales must be modeled and removed \citepalias{M87PaperIII}.
Along the frequency axis, residual post-CALC
delays---phase slopes over the frequency band---are caused primarily by atmospheric path-length variations.
No significant instrumental delay effects are present in our digital recording system.
However, imperfect amplitude- and phase-bandpass responses impact the data and must be corrected.

We have developed two independent \ac{vlbi} data reduction pipelines to perform these pre-averaging calibration steps to stabilize the signal: EHT-HOPS \citep{2019ApJ...882...23B}, which is based on the \ac{hops} \citep{2004RaSc...39.1007W} software and rPICARD \citep{2018evn..confE..80J, 2019A&A...626A..75J}, which is based on the \ac{casa} \citep{2007ASPC..376..127M, 2019arXiv190411747V} package. The other papers in this series make use of the data produced from both pipelines, whose consistency has been established in \citetalias{M87PaperIII}, for a verification of scientific results. The AIPS-based \citep{2003Greisen} pipeline that has also been used for the \m87 EHT results \citepalias{M87PaperIII} is no longer being maintained.

EHT-HOPS processes data that have been converted into Mark4 format by DiFX task \texttt{difx2mark4}. These correlation coefficients have been normalized to unity autocorrelation at 0.4\,s $\times$ 58\,MHz resolution and then scaled to idealized analog correlation amplitudes according to the 2-bit quantization efficiency correction factor $\sim$1/0.88.
Baseline rate and delay solutions are fit with the \ac{hops} \texttt{fourfit} routine. These fringe solutions are then ``globalized'' into station-based corrections using a least-squares method \citep[similar to][]{alef1986}. A stable phase bandpass response for each antenna is derived using an ensemble of high \ac{sn} detections on bright calibrators and applied to all data.
Turbulent phases introduced by the troposphere are corrected by fitting a piecewise polynomial phase model to the visibilities from baselines connected to the most sensitive station in each scan.
To avoid overfitting to thermal noise, the phases of each 58\,MHz spectral window are independently corrected using a model derived from the other, remaining 31 spectral windows.
Finally, the geometric feed rotation angle evolution is corrected and relative complex gains between the RCP and LCP signal paths are fitted.
We commonly refer to the calibrated data produced by the EHT-HOPS pipeline as ``HOPS'' data.

rPICARD follows the Hamaker-Bregman-Sault measurement equation \citep{1996Hamaker,oms2011, oms2011a, oms2011b, oms2011c} as implemented in the \ac{casa} framework.
For the amplitude calibration, the \ac{casa} \texttt{accor} task is used to enforce unity auto-correlations, correcting for digital sampler biases. As this task scales the visibilities at the 58\,MHz resolution, it also corrects for the gross amplitude bandpass of each station.
Residual amplitude-bandpass effects are removed with a custom bandpass calibration table formed by taking the median of the normalized 0.5\,MHz channelized auto-correlations of all \ac{vlbi} scans combined.
The phases are corrected with the \texttt{fringefit} task, which performs a global \citet{Schwab1983} fringe-fit assuming an unpolarized
point source model.
The data are segmented into the shortest bins within the expected atmospheric coherence time, where a sufficiently high \ac{sn} can be accumulated to obtain fringe detections. These detections are then used to correct for atmospheric phase turbulence and small residual delay variations that can occur within \ac{vlbi} scans.
A crude phase bandpass at 58\,MHz is solved with a ``single-band fringe-fit'', where we apply the solutions from the scan with the highest \ac{sn} fringe solution per antenna.
The residual 0.5\,MHz phase bandpass is corrected with the \ac{casa} \texttt{bandpass} task with a \ac{sn}\,$>$\,3 cutoff using the data of all calibrator scans combined.
Geometric feed rotation angles are corrected for on-the-fly, following the measurement equation, and no relative complex gains are corrected in the \ac{casa} data.
We commonly refer to the calibrated data produced by the rPICARD pipeline as ``CASA'' data.

The EHT \ac{vlbi} data indicate a phase offset (either constant, or with a time-dependent drift component) between the two polarization channels that can be attributed largely to instrumental effects, and in a small part to circular polarization (Stokes $\mathcal{V}$) of the source. The instrumental phase shifts are corrected through polarimetric gain ratio calibration, aligning RR and LL components to compute the total-intensity (Stokes $\mathcal{I}$) visibilities coherently.
Leveraging the facts that the RCP$-$LCP instrumental phase of \ac{aa} after PolConvert is zero \citep{martividal2016, 2019PASP..131g5003G} and that intrinsic Stokes $\mathcal{V}$ signals will have a negligible phase contributions on \ac{aa} baselines, the parralel hand signals have been aligned for each \ac{vlbi} scan under the assumption of $\mathcal{V}=0$ on \ac{aa} baselines.
The same strategy is also employed inherently by rPICARD, as the parallel correlation products are fringe-fitted separately. As the phase calibration is station-based, intrinsic Stokes $\mathcal{V}$ signals are not removed from the data.
%
More details about the EHT data reduction pipelines, including flow-charts of their processing steps, and detailed verification tests are given in \citetalias{M87PaperIII}.\footnote{Note that the \m87 HOPS data presented in \citetalias{M87PaperIII} were obtained with multi-source, multi-day calibration of the polarimetric gain ratios \citep{Steel2019}. While the aim of this method was to preserve the Stokes $\mathcal{V}$ structure of the resolved images, it indicated underfitting of the instrumental effects, potentially increasing the resulting systematic errors.}

Due to the excellent sensitivity of the \ac{eht}, we have a high \ac{sn} on many baselines probing \ac{sgra}, which allows us to model atmospheric phase fluctuations occurring on short timescales to effectively extend the coherence time \citepalias{M87PaperIII}. 
Combined with the small field of view around the \ac{vlbi} correlation phase center of the \ac{eht}, this means that reasonable averaging intervals are limited primarily by the spectral and temporal variations of the target.
With a fractional bandwidth of $\sim 1\,\%$, we can safely average the data over our observing bandpass without introducing considerable bandwidth smearing effects \citep{TMS}.
On the other hand, \ac{sgra} is known to vary on very short timescales, possibly down to the gravitational timescale of $\sim 21$\,s.
We have therefore chosen to average our data, which was correlated with a 0.4\,s period, into 10\,s time bins.


\subsection{Flux density scaling}
\label{sec:calib_fluxcal}

\begin{table}
\caption{Fitted LMT gains on each EHT observing day.}
\setlength{\tabcolsep}{4pt}
\label{tab:lmt-gain}
\begin{center}
\begin{tabularx}{1.0\linewidth}{@{\extracolsep{\fill}}ccc}
\hline
\hline
UTC Day      & RCP DPFU    & LCP DPFU \\
(April 2017) & (Jy/K) & (Jy/K)\\
\hline
05   & $0.046 \pm 0.003$ &   $0.048 \pm 0.003$ \\ 
06   & $0.032 \pm 0.009$ &   $0.034 \pm 0.009$ \\ 
07   & $0.061 \pm 0.001$ &   $0.064 \pm 0.001$ \\ 
10   & $0.076 \pm 0.001$ &   $0.077 \pm 0.001$ \\ 
11   & $0.067 \pm 0.005$ &   $0.067 \pm 0.004$ \\ 
\hline
\end{tabularx}
\end{center}
\end{table}

To scale our correlation coefficient measurements to a physical flux density unit scale, we estimate the \ac{sefd} of every station in the array. The \ac{sefd} of a single antenna or phased-up array is given as
\begin{equation}
    \mathrm{SEFD} = \frac{T_\mathrm{sys}^*}{\eta_\mathrm{ph} \, \mathrm{gc}(E) \, \mathrm{DPFU}} \, .
\end{equation}
Here, $T_\mathrm{sys}^*$ is the effective system temperature, which characterizes the total noise contribution along a station's signal path and corrects for the signal attenuation caused by the Earth's atmosphere.
The phasing efficiency of a phased array is given by $\eta_\mathrm{ph}$, which is unity for single dish stations.
The station gain is factored into a normalized elevation ($E$) gain curve $\mathrm{gc}(E)$ and the ``degrees-per-flux-unit'' (DPFU) conversion factor between measured system temperatures and flux densities.
The DPFU is determined by the aperture efficiency $\eta_\mathrm{ap}$ and total collecting area $A_\mathrm{geom}$ as $\mathrm{DPFU} = \eta_\mathrm{ap} A_\mathrm{geom} / (2 k_\mathrm{B})$, with $k_\mathrm{B}$ the Boltzmann constant.
We can scale a correlation coefficient $r_{i-j}$ measured on a baseline $i$\,--\,$j$ in units of thermal noise, to a visibility $V_{i-j}$ in units of Jy via
\begin{equation}
    V_{i-j} = \sqrt{\mathrm{SEFD}_i \mathrm{SEFD}_j} r_{i-j} \,.
\end{equation}

\begin{figure}
\includegraphics[width=\columnwidth]{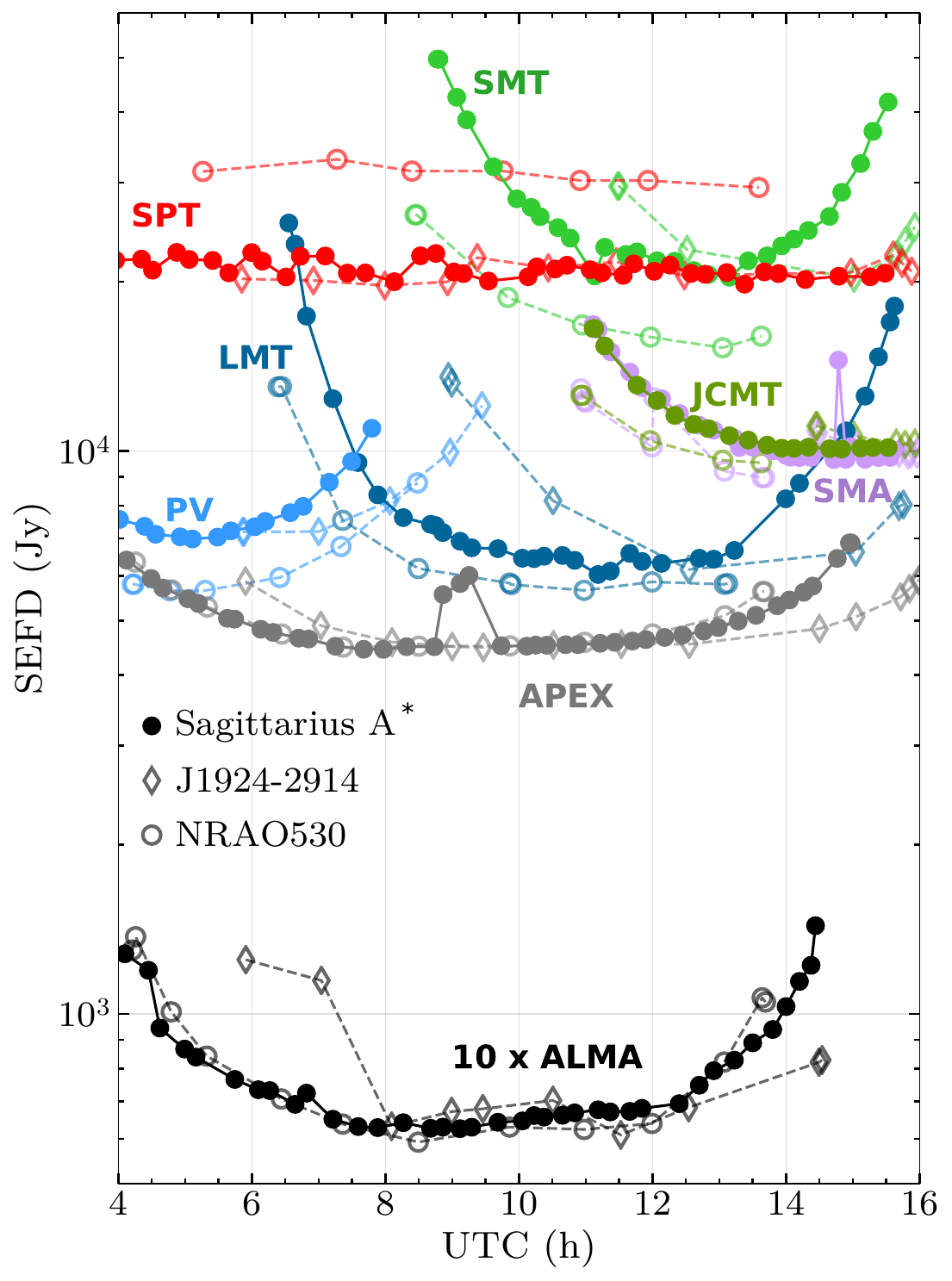}
\caption{Example of system equivalent flux density (SEFD) values during a single night of the 2017 EHT observations (2017 April 7, high-band RCP). Values for \ac{sgra} are marked with full circles, values for J1924$-$2914 are marked with empty diamonds, and values for NRAO\,530 are marked with empty circles. ALMA SEFDs have been multiplied by 10 in this plot for visualization purposes.}
\label{fig:sefd}
\end{figure}

We have performed this flux density calibration with a common framework\footnote{\url{https://github.com/sao-eht/eat}} for both the \ac{casa} and \ac{hops} data.
Compared to the \acp{sefd} used to calibrate the data released for \m87 and \3c279 \citepalias{M87PaperIII}, we have updated several a priori calibration parameters.
We are using the \citet{Butler-planet-models} planet models (updated for the year 2017) for the gain calibration of the \ac{ap}, \ac{pv}, and \ac{az} telescopes, instead of the previously used GILDAS\footnote{\url{http://www.iram.fr/IRAMFR/GILDAS}} models.
Additionally, we have used \ang{;;10.1} instead of \ang{;;8.5} as an updated estimate for the beam size of the LMT and subsequently fitted for the variable station gain on a per-day basis (\autoref{tab:lmt-gain}).

Data from SMA were re-analyzed, using the intra-SMA interferometric data to derive new \ac{sefd} estimates for the phased array data. Aperture efficiencies were calculated using observations of the solar system objects Callisto, Ganymede, and Titan, using the brightness temperature models from \citet{Butler-planet-models}, and derived separately for low and high band. Amplitude gain solutions were derived using two quasars---NRAO\,530 and J1924$-$2914---that were observed periodically during the observations of \sgra. Phase-only self-calibration was applied to the \ac{sm} interferometer data of \sgra, assuming a point-source model and excluding baselines shorter than 15 k$\lambda$ in length, due to the presence of large-scale emission features. The resultant gain solutions were used to re-derive phasing efficiencies for the beamformed data used for \ac{vlbi} analysis. More details about the \ac{sm} data reduction are given in \citet{Wielgus2022} Section 2.3.

Typical \ac{sefd}s used to calibrate the \sgra flux density are shown in \autoref{fig:sefd}. For the time-averaged spectral energy distribution (SED) data gathered in Table~\ref{tab:sed_2017}, we calculate a flux density of $2.4\pm0.2$ Jy between $213-229$ GHz.

\begin{table*}
\caption{Values for the \sgra SED observed by coordinated ground-based and space-based observatories during the EHT 2017 run.}             
\label{tab:sed_2017}      
\begin{tabularx}{\linewidth}{@{\extracolsep{\fill}}l c c c c}
\hline\hline
Observatory & Frequency & $\nu$F$_{\nu}$ & Flux & $\nu$L$_{\nu}$ \\ 
&[GHz]&[$\times10^{-12}$erg\,s$^{-1}$\,cm$^{-2}$]&[Jy]&[$\times10^{34}$erg\,s$^{-1}$]\\
\hline  
   EAVN & 22 & 0.24 $\pm$ 0.02 & 1.07 $\pm$ 0.11 & 0.19 $\pm$ 0.02 \\
   EAVN & 43 & 0.58 $\pm$ 0.06 & 1.35 $\pm$ 0.14 & 0.48 $\pm$ 0.05 \\
   GMVA & 86 & 1.63 $\pm$ 0.17 & 1.9 $\pm$ 0.2 & 1.3 $\pm$ 0.1 \\
   GMVA & 88 & 1.67 $\pm$ 0.18 & 1.9 $\pm$ 0.2 & 1.4 $\pm$ 0.1 \\
   GMVA & 98 & 1.96 $\pm$ 0.20 & 2.0 $\pm$ 0.2 & 1.6 $\pm$ 0.2 \\
   GMVA & 100 & 2.10 $\pm$ 0.20 & 2.1 $\pm$ 0.2 & 1.67 $\pm$ 0.2 \\
   ALMA/SMA\textsuperscript{a} & 213-229 & 5.28 $\pm$ 0.44 & 2.4 $\pm$ 0.2 & 4.3 $\pm$ 0.4 \\
   VLT & 1.38$\times10^{5}$ & $<$4.12 & $<$0.003 & $<$3.4 \\
   Chandra/NuSTAR & 5.68$\times10^{8}$ & $0.70_{-0.20}^{+0.15}$ & $1.23_{-0.35}^{+0.27}  \times10^{-7}$ & $0.57_{-0.16}^{+0.13}$ \\
   Chandra/NuSTAR & 7.84$\times10^{8}$ & $0.49_{-0.11}^{+0.06}$ & $6.27_{-1.40}^{+0.74} \times10^{-8}$ & $0.40_{-0.09}^{+0.05}$ \\
   Chandra/NuSTAR & 1.08$\times10^{9}$ & $0.33_{-0.05}^{+0.01}$ & $3.09_{-0.46}^{+0.11} \times10^{-8}$ & $0.27_{-0.04}^{+0.01}$ \\
   Chandra/NuSTAR & 1.49$\times10^{9}$ & $0.25_{-0.04}^{+0.25}$ & $1.71_{-0.25}^{+1.71} \times10^{-8}$ & $0.21_{-0.03}^{+0.21}$ \\
   Chandra/NuSTAR & 2.06$\times10^{9}$ & $0.08_{-0.02}^{+0.09}$ & $0.39_{-0.11}^{+0.43} \times10^{-8}$ & $0.06_{-0.02}^{+0.07}$ \\
\hline
   Chandra\textsuperscript{b} & 1.57$\times10^{9}$ & $5.63_{-3.36}^{+5.06}$ & $3.6_{-2.1}^{+3.2} \times10^{-7}$ & $4.61_{-2.76}^{+4.15}$ \\
   NuSTAR\textsuperscript{b} & 9.55$\times10^{9}$ & $0.78_{-0.73}^{+0.86}$ & $8.3_{-7.6}^{+9.0} \times10^{-9}$ & $0.64_{-0.60}^{+0.70}$ \\
\hline
\end{tabularx}
\newline
\textsuperscript{a}\footnotesize{Mean measurement across 213-229 GHz. The spectral index at these frequencies was observed to be close to zero \citep{Wielgus2022}} \\
\textsuperscript{b}\footnotesize{2017 April 11 X-ray flare.}
\tablecomments{Frequencies for X-ray observatories reflect the central frequency of the keV energy band within the observation bin.}
\end{table*}

\subsubsection{Light curves of \sgra and ALMA-VLBI amplitude calibration}
\label{sub:minispiral_cal}


The ALMA Phasing System is designed to construct a summed, formatted \ac{vlbi} signal without interrupting the data stream of each individual antenna, thus allowing the ALMA correlator to compute all of the intra-ALMA visibilities at the same time \citep{2018PASP..130a5002M}.
The calibration information related to these intra-ALMA visibilities is needed for the correct polarimetric processing of the ALMA-VLBI signal \citep{martividal2016,2019PASP..131g5003G, 2021Goddi}.

At the spatial scales sampled by the intra-ALMA baselines, the \sgra field consists of the sum of two components: a point-like source located at the field center and with a time-dependent flux density (i.e., the AGN), and an extended structure that covers several arcseconds, with a total flux of about 1.1\,Jy and a surface brightness of about 0.12\,Jy/beam \citep{Wielgus2022}. This extended component is known as the ``minispiral'', and its emission is related to ionized gas and dust in the galactic center \citep[e.g.,][]{Lo1983}.

The use of the intra-ALMA self-calibration gains to compute the corrections for the VLBI amplitudes of the ALMA-related baselines provides a very accurate determination of the relative changes in the ALMA's VLBI amplitude gains \citep{2019PASP..131g5003G}.  
However, this approach has important limitations for the correct calibration of a time-variable source like \sgra (\autoref{fig:mwl_cover}). In particular, the use of an a priori (constant) model for the amplitude self-calibration of the \sgra intra-ALMA visibilities results in an incorrect estimate of the phased-ALMA amplitude scaling. 
To overcome this limitation, we calibrate the \sgra \ac{aa} gains with the source's light curve, which is computed under the assumption that the flux density distribution of the minispiral remains stable across the extent of the observations \citep{Wielgus2022}.

We note that, even though the minispiral modeling allows us to calibrate relative \ac{aa} gains at the $\sim$\,1\,\% level, a constant $\sim$\,10\,\% calibration uncertainty remains for the overall gain of the phased array, which is tied to the estimated flux density of Ganymede \citep[the source used as primary calibrator,][]{2019PASP..131g5003G}.

We perform a time-dependent variant of network calibration \citepalias{M87PaperIII} of the EHT VLBI data to a merged light curve from \ac{aa} and the \ac{sm} (\autoref{fig:mwl_cover}). The \ac{aa} light curve is computed through the minispiral method described above and a description of the SMA light curve is given in \citet{Wielgus2022}. The two light curves are combined by leveling the median amplitudes in the overlapping time between ALMA and SMA and a smoothing spline interpolation in time. The network calibration procedure leverages the presence of co-located sites, constraining gains of telescopes with a co-located partner by assuming that intra-site baselines (ALMA--APEX and JCMT--SMA; see \autoref{fig:schedules}) observe a point source with a time-dependent flux density, corresponding to the light curve.

\section{Multi-wavelength Observing Campaign}
\label{sec:mwl_obs}

In addition to the ALMA and SMA  millimeter light curves collected as a part of the EHT observations \citep{Wielgus2022}, the 2017 \sgra EHT campaign includes observations from elite ground-based facilities (\S \ref{subs:ground_obs}), including 
the East Asian VLBI Network, the Global 3\,mm VLBI Array, the Very Large Telescope, as well as space-based telescopes (\S \ref{subs:space_obs}) including the Neil Gehrels Swift Observatory, the Chandra X-ray Observatory, and the Nuclear Spectroscopic Telescope. 
Figure \ref{fig:mwl-coverage} shows coverage of \sgra for each of these instruments during the campaign. 
These coordinated observations provide (quasi-)simultaneous wavelength coverage 
and enable detailed multi-wavelength variability studies that place broadband constraints on models (see \citetalias{PaperV} and \citetalias{PaperVI}).
We describe these observations briefly here and place them in the broader EHT and \sgra historical context in Section \ref{subs:sup_data}.

\subsection{Supplementary Ground-based Observations}
\label{subs:ground_obs}

\subsubsection{East Asian VLBI Network}
\label{subsubs:eavn_obs}

The East Asian VLBI Network \citep[EAVN; e.g.,][]{Wajima_2016, An_2018, Cui2021} consists of the 7 telescopes of KaVA (KVN\footnote{Korean VLBI Network: three 21\,m telescopes in Korea (Yonsei, Ulsan, and Tamna)} and VERA\footnote{VLBI Exploration of Radio Astrometry: four 20\,m telescopes in Japan (Mizusawa, Iriki, Ogasawara, and Ishigakijima)} Array; e.g., \citealt{Lee_2014, Niinuma_2015}), and additional telescopes of the Japanese VLBI Network \citep[JVN; e.g.,][]{Doi_2006} and the Chinese VLBI Network \citep[CVN; e.g.,][]{Zheng_2015}. 
%
%
During the EHT 2017 window, four EAVN observations were carried out at 22 and 43\,GHz (\autoref{fig:mwl-coverage}). A single, symmetric Gaussian model was found to describe the intrinsic structure of \ac{sgra} at both wavelengths. The measured flux densities from \sgra at these frequencies are $(1.07\pm0.11)$\,Jy and $(1.35\pm0.14)$\,Jy, respectively (Table \ref{tab:sed_2017}).
Two of the  on 2017 April 3 and 4 are (quasi-)simultaneous with the Global 3\,mm VLBI Array observations (Section~\ref{subsubs:gmva_obs}) as well as the EHT sessions \citep{Cho_2022}.
These measurements provide an estimated size and flux density of \sgra at 1.3\,mm via extrapolation of power-law models (i.e., the intrinsic size scales with observing wavelength as a power-law with an index of $\sim1.2\pm0.2$).


\subsubsection{Global 3\,mm VLBI Array}
\label{subsubs:gmva_obs}


VLBI observations of \sgra at 86\,GHz were conducted on 2017 April 3 with the Global Millimeter VLBI Array (GMVA)\footnote{\href{https://www3.mpifr-bonn.mpg.de/div/vlbi/globalmm/}{https://www3.mpifr-bonn.mpg.de/div/vlbi/globalmm/}}.
Eight Very Long Baseline Array antennas equipped
with 86\,GHz receivers, the Robert C. Byrd Green Bank Telescope (GBT), the Yebes 40\,m
telescope, the Effelsberg
100\,m telescope, \ac{pv}, and 37 phased \ac{aa} antennas participated in the observation (project code MB007, published in \citealt{2019ApJ...871...30I}).

The data were recorded with a bandwidth of 256\,MHz for each polarization and fringes were detected out to 2.3\,G$\lambda$. The total on-source integration time on \sgra was 5.76\,hr over a 12\,hr track, with ALMA co-observing for 8\,hr.
The results of the experiment rule out jet-dominated radio emission models of \sgra with large viewing angles ($>20\deg$) and provide stringent constraints on the amount of refractive noise added by the interstellar scattering screen towards the source \citep{2019ApJ...871...30I}, discussed in more detail in Section~\ref{sec:scattering}. The total flux density measured at 86\,GHz is $(1.9 \pm 0.2)$\,Jy 
(Table \ref{tab:sed_2017}). 

\subsubsection{Very Large Telescope}
\label{subsubs:vlt_obs}

The Paranal Observatory's Nasmyth Adaptive Optics System (NAOS) and Near-Infrared Imager and Spectrograph (CONICA) instrument on the Very Large Telescope (VLT), also known as VLT/NACO, measured a K-band near infrared 
upper limit of 3\,mJy during the 2017 April 7 EHT observing run (courtesy of the MPE Galactic Center Team). 
Ongoing observations with the new Very Large Telescope Interferometer GRAVITY instrument \citep[VLTI/GRAVITY,][]{2017A&A...602A..94G} indicate that \sgra's typical flux distribution in the NIR K-band changes slope at a median flux density of $1.1\pm0.3$ mJy,
characteristic of \sgra's quiescent NIR emission \citep{GRAVITY2020}.  

\subsection{Coordinated Space-based Observations} 
\label{subs:space_obs}


\subsubsection{Neil Gehrels Swift Observatory}
\label{subs:swfit_obs}

%
%
%
%
%
%
Observations from the \swift\ 
X-ray Telescope \citep[XRT,][]{Gehrels2004, Burrows2005} were reprocessed with the latest calibration database files and the \swift\ tools contained in HEASOFT-v6.20\footnote{\url{https://heasarc.gsfc.nasa.gov/lheasoft/}}. Source flux in the 2--10 keV energy band is extracted from a 10\arcsec\ radius circular region centred on the position of \sgra.  
Count rates are reported as measured \citep[e.g.,][]{Degenaar13}, i.e., without any correction for the known significant absorption along the \sgra\ sightline ($\rm N_H \sim 9\times10^{22}\,cm^{-2}$).

There are 48 \swift\ observations of the Galactic center between 2017 April 5 and 2017 April 12, with a total exposure time of 26.3 ks (\autoref{fig:mwl-coverage}). These observations include a dedicated dense sampling schedule to coincide with the EHT observing window and two observations from the regular Galactic center monitoring program \citep{Degenaar2013,Degenaar2015,vdEijnden2021}. The average exposure time of the dense sampling was $\sim 500$\,s, with an average interval between observations of $\sim 3.5$\,hr.

In the 2017 April 7 \swift\ observation that overlaps the EHT window (\autoref{fig:mwl_cover}), a 2--10 keV flux is detected (0.023\,ct\,s$^{-1}$) in excess of the 2017 2$\sigma$ trend-line (0.018\,ct\,s$^{-1}$), as measured from the cumulative flux distribution observed from \sgra.
None of the \swift\ observations are simultaneous with the 2017 April 11 \chandra flare described in the following section. 

\subsubsection{Chandra X-ray Observatory}
\label{subs:chandra_obs}

A series of Chandra X-ray Observatory \citep{Weisskopf2002} exposures of \sgra\ were acquired on 2017 April 6, 7, 11, and 12 using the ACIS-S3 chip in FAINT mode with a 1/8 subarray (observations IDs 19726, 19727, 20041, 20040; PI: Garmire), for a total of $\sim$133 ks coordinated with the EHT campaign (\autoref{fig:mwl-coverage}). The small subarray mitigates photon pileup during bright \sgra\ flares, as well as contamination from the magnetar SGR J1745$-$2900, which peaked at X-ray wavelengths in 2013 and has faded over the more than 6 years since \citep{2013ApJ...770L..23M, 2013ApJ...775L..34R, 2020ApJ...894..159R, 2015MNRAS.449.2685C, 2017MNRAS.471.1819C}. It also achieves a frame rate of 0.44\,s vs.\ \chandra's standard rate of 3.2\,s.

\begin{figure*}
\includegraphics[width=\textwidth]{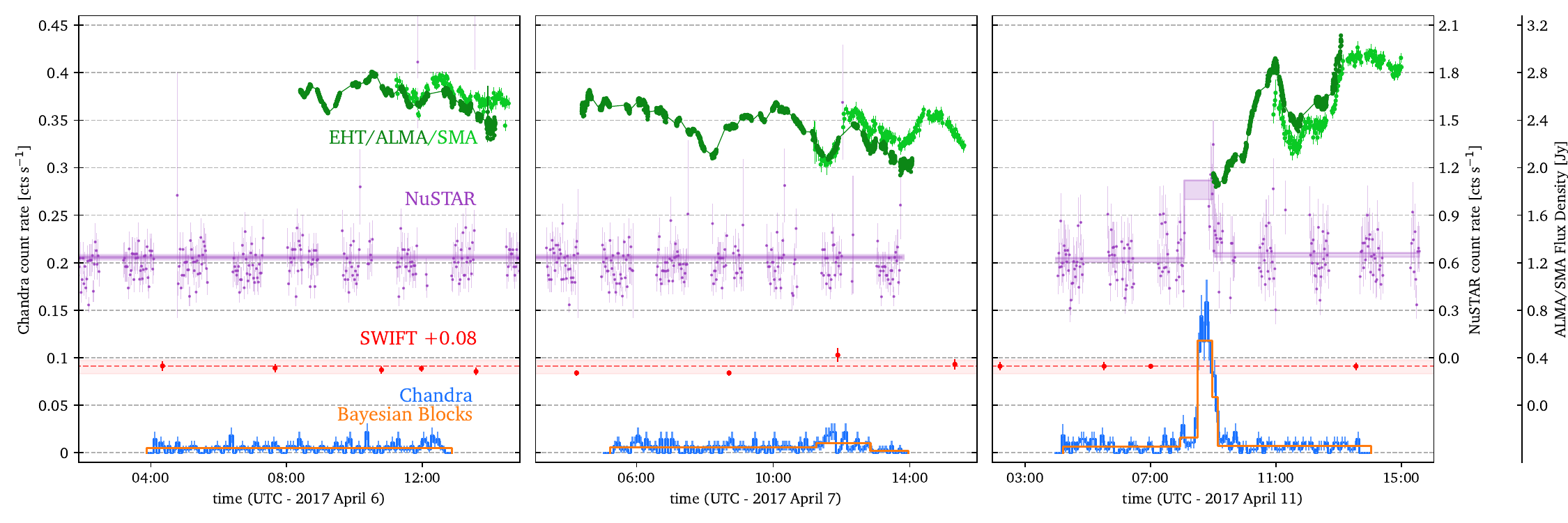}
\caption{
Light curves from ALMA and SMA between $213-229$ GHz \citep[dark and light green points, respectively; see][for details]{Wielgus2022}, along with (quasi-)simultaneous, coordinated \swift\ (red; 2--10 keV), \chandra\ (blue; 2-–8 keV), and \nustar (purple; 3--79 keV) observations on 2017 April 6, 7, and 11. The long term average flux value of \sgra measured from \swift\ is plotted as a red dashed line with upper and lower $2\sigma$ percentiles for the 2017 cumulative flux distribution marked with a light red shaded region \citep[e.g.,][]{Degenaar2013, Degenaar2015}. The Bayesian block flare detection algorithm \citep{Scargle1998,Scargle2013,Williams2017} has been run on the \chandra\ and \nustar\ light curves and results, including flare detections on 7 and 11 April, are over-plotted as orange histograms on the \chandra\ data and purple histograms on the \nustar\ data. 
\label{fig:mwl_cover}}
\end{figure*}

\chandra\ data reduction and analysis are performed with the CIAO v4.13 package{\footnote{Chandra Interactive Analysis of Observations (CIAO) is available at \url{http://cxc.harvard.edu/ciao/}.}} \citep{2006SPIE.6270E..1VF}, CALDB v4.9.4. We use the \texttt{chandra\_repro} script to reprocess the level 2 events files, update the WCS coordinate system (\texttt{wcs\_update}), and apply barycentric corrections to the event times (\texttt{axbary}). The $2-8$ keV light curves are then extracted from a circular region of radius 1.25\arcsec\, centered on the radio position of \sgra. 
Light curves for 2017 April 6, 7 and 11 are shown in \autoref{fig:mwl_cover}. Using the Bayesian Blocks algorithm \citep{Scargle1998,Scargle2013,Williams2017}, we search these light curves for flares and robustly detect one on 2017 April 11, with a second weaker detection on April 7 (orange histograms overplotted on the \chandra light curves in Fig. \ref{fig:mwl_cover}).

We use {\tt specextract} to extract X-ray spectra and response files from a similar 1.25\arcsec\ region, centered on \sgra.  
Since our primary interest for this dataset is the flare emission, we do not extract background spectra from a separate spatial region. Instead, spectra of the quiescent off-flare intervals play the role of our background spectra.

The flare and off-flare intervals are identified by analyzing the X-ray light curves of each observation. For the easily detectable flare on 2017 April 11 (observation ID 20041), we use the direct Gaussian fitting method presented in \citet{Neilsen2013b}. For 2017 April 7 (observation ID 19727), we use the Bayesian Blocks decomposition \citep{Scargle1998,Scargle2013,Williams2017}; 
this method is better suited to detecting the sustained low-level activity apparent toward the end of the observation.






\subsubsection{Nuclear Spectroscopic Telescope Array}
\label{subs:nustar_obs}

X-ray observations from the Nuclear Spectroscopic Telescope Array \cite[\nustar][]{Harrison2013} 
performed three \sgra observations from 2017 April 6 to 2017 April 11 (observation IDs: 30302006002, 30302006004, 30302006006). These 
provide a total exposure time of $\sim103.9$\,ks and were coordinated with the EHT campaign. We reduced the data using the \nustar Data Analysis Software NuSTARDAS-v.1.6.0\footnote{\url{https://heasarc.gsfc.nasa.gov/docs/nustar/analysis}.} and HEASOFT-v.6.19, filtered for periods of high instrumental background due to South Atlantic Anomaly (SAA) passages and known bad detector pixels. Photon arrival times were corrected for on-board clock drift and processed to the Solar System barycenter using the JPL-DE200 ephemeris. We used a source extraction region with 50\arcsec\ radius centered on the radio position of \sgra\ and 
extracted 3--79 keV light curves in 100\,s bins with deadtime, PSF, and vignetting effects corrected \citep[see][for further details on \nustar \sgra data reduction]{Zhang17}. For all three observations we made use of the data obtained by both focal plane modules FPMA and FPMB.

For flare spectral analysis, we used {\tt nuproducts} in HEASOFT-v.6.19 to create spectra and responses from 30\arcsec\ circular regions (as recommended for faint sources to minimize the background) centered on the coordinates of \sgra.  
As explained in \ref{sec:joint-xray}, \nustar only partially detected a flare from \sgra on 2017 April 11 (observation ID 30302006004), hence we focus on this observation alone in the present work. The source spectrum was extracted from the \nustar Good Time Intervals (GTIS) that overlaps with the \chandra flare duration. The background spectrum was extracted from
off-flare time intervals in the same observation. 

\begin{figure}[h]
    \includegraphics[width=\columnwidth]{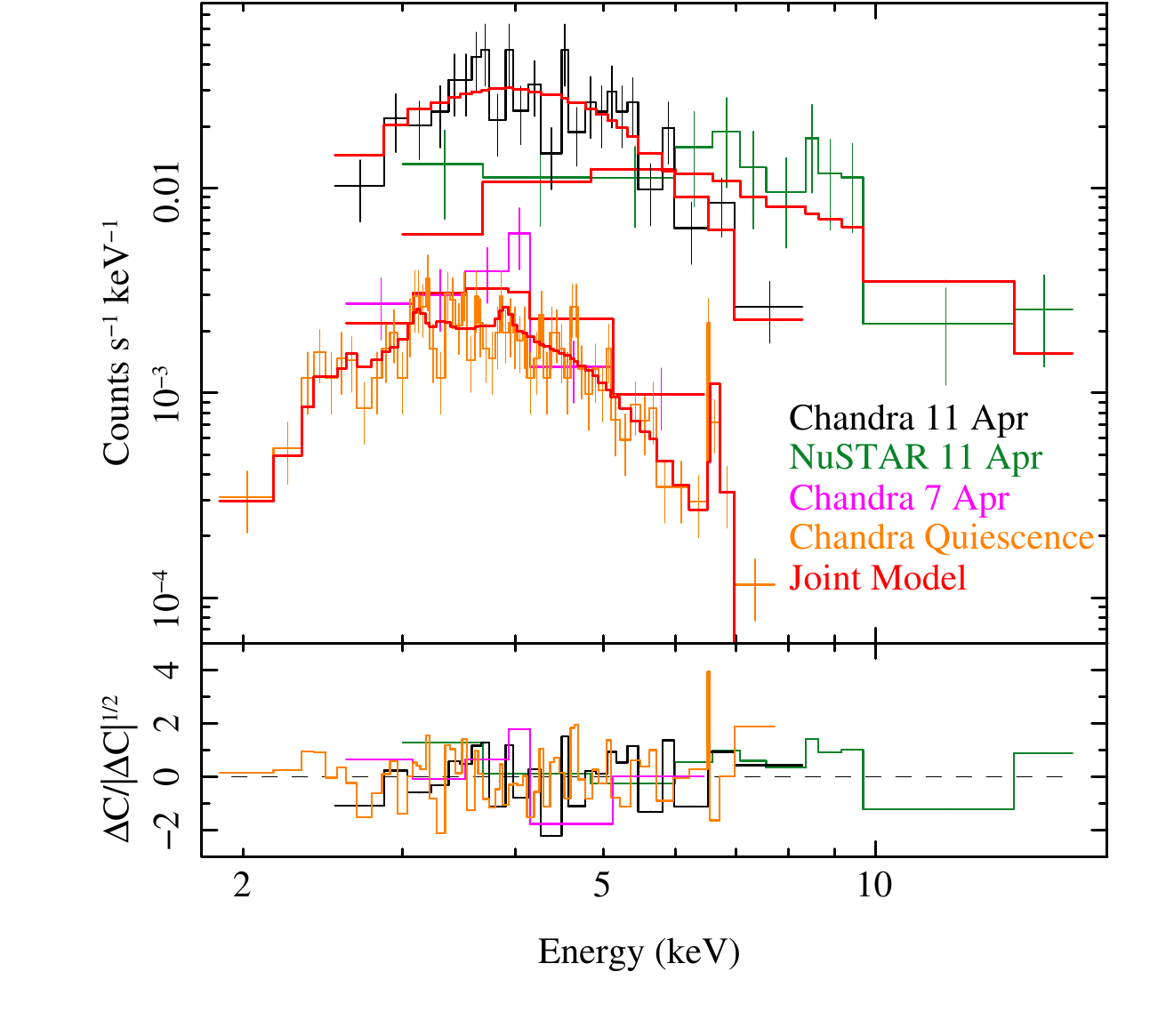}
\caption{Joint \chandra-\nustar spectra of flares from \sgra. The magenta and black spectra are the \chandra flare spectra from 2017 April 7 and 2017  April 11, respectively; the 2017 April 11 \nustar FPMA and FPMB flare spectra are combined for plotting purposes and shown in green. The orange data are the combined spectrum from the off-flare quiescent intervals of all four \chandra observations. The model described in Section \ref{sec:joint-xray} is shown in red for each dataset.}
\label{fig:Xrayspectra}
\end{figure}

\subsubsection{X-ray Flare Spectral Analysis}
\label{sec:joint-xray}
To obtain the best constraint on the spectrum of \sgra during its faint and moderate X-ray flares, we performed joint \chandra/\nustar spectral analysis of the two flares described in Section \ref{subs:chandra_obs}. The following analysis was performed in ISIS v1.6.2-43 and made use of Remeis {\tt isisscripts}
\footnote{This research has made use of a collection of ISIS functions (ISISscripts) provided by ECAP/Remeis observatory and MIT (\url{http://www.sternwarte.uni-erlangen.de/isis})}. \chandra flare spectra were binned to a minimum of four channels and a \ac{sn} ratio of 3 above 0.5\,keV while off-flare spectra were combined and binned to a minimum of two channels and \ac{sn} of 3; we fit bins contained within the interval 1--9\,keV. The \nustar FPMA/FPMB spectra of the 2017 April 11 flare were binned to a combined minimum  \ac{sn} of 2 and a minimum of three channels above 3\,keV. We ignored all bins not fully contained within the interval 3--79\,keV.

Given the relatively small numbers of counts, we opted for simple absorbed power law models for the flare spectra and we compared our model to our data with Cash statistics \citep{Cash1979}. We used the {\tt TBvarabs} model with Wilms abundances and Verner cross-sections \citep{Verner1996, Wilms2000} and assumed a shared spectral index for both flares. (There is no conclusive evidence for a relationship between X-ray flare luminosity and X-ray spectral index; \citealt{Neilsen2013b,Zhang2017,Haggard2019}.) Because a portion of the 2017 April 11 flare fell within a gap in the \nustar light curve, we included a cross-normalization constant between our \nustar and \chandra spectra (as well as between the FPMA and FPMB spectra).  

For \nustar, we simply defined the spectrum of the off-flare interval as the background file for the 2017 April 11 flare spectrum, but we had to treat the \chandra spectra differently because they are susceptible to pileup during flares \citep[e.g.,][]{Nowak2012}, though fainter flares like those described here are not likely to be impacted. In particular, because pileup depends on the total count rate, not the background count rate, it is necessary to model the background spectrum and treat the flare emission as the sum of the background model and the absorbed power law. We fit the quiescent (off-flare) emission with a single {\tt vapec} model; see \citet{Nowak2012} for more details.

Once we had a satisfactory fit to the joint flare spectra, we used the {\tt emcee} Markov Chain Monte Carlo (MCMC) routine to find credible intervals for our fit parameters. For our MCMC runs, we used 10 walkers (i.e., 10 members of the ensemble) 
for each of our 10 free parameters, and allowed them to evolve for 10,000 steps (a total of 1 million samples). These runs appear to converge within the first several tens of steps, so we discard the first 500 steps. We estimate an autocorrelation time for our parameter chains between $\sim200$ and $400$ steps, indicating that we have 2500--5000 independent samples of each parameter.

Finally, we calculate the minimum width 90\% credible interval for each parameter. We compute the cumulative distribution function for the samples for each parameter and select the smallest interval that contains 90\% of the samples. The results are given in Table \ref{tab:xrayflares}. 
%
\ac{sgra} flares are highly absorbed, with a column density of $N_{\rm H}=17.8_{-2.5}^{+3.5}$\,cm$^{-2};$ this is a bit higher than the values found by \citet{Nowak2012} and \citet{Wang2013}, but the differences are within 1-$\sigma$ in both cases. 
The photon index of the flare is $\Gamma=2.1_{-0.4}^{+0.5}$, which is not well constrained but is consistent with other analyses of X-ray flares \citep{Porquet2008,Nowak2012,Neilsen2013,Haggard2019}. The 2017 April 7 
flare is only detected at 99\% confidence, 
but has an unabsorbed 2--10\,keV flux of $F_{\rm 0704,2-10}$ = $(0.3\pm0.2\times10^{-12}$)\,erg\,s$^{-1}$\,cm$^{-2}$ and a  3--79\,keV flux $F_{\rm 0704,3-79} =(0.5_{-0.4}^{+0.8})\times10^{-12}$\,erg\,s$^{-1}$\,cm$^{-2}$. The 2017 April 11 flare has an unabsorbed 2--10\,keV flux of $F_{\rm 1104,2-10}$ = $(7.8_{-3.9}^{+2.6})\times10^{-12}$\,erg\,s$^{-1}$\,cm$^{-2},$ which rises to $F_{\rm 1104,3-79} =(15.4_{-7.5}^{+8.9})\times10^{-12}$\,erg\,s$^{-1}$\,cm$^{-2}$  over the interval 3--79\,keV. 

As reported in \citet{Neilsen15}, the quiescent X-ray flux ($F_{\rm Q,2-10}$) results from an admixture of two components. At the faint end of \sgra's flux distribution, there are both variable and steady components, with the variable component likely arising from unresolved faint flares which contribute $\sim$10\% of the apparent quiescent flux \citep[see also][]{Neilsen13}. Adopting the $2-10$ keV flux from this joint fit (Table \ref{tab:xrayflares}) and allowing a generous $\sim 1/3$ contribution 
from unresolved X-ray flares, 
we estimate an upper limit on the median luminosity of $\sim 10^{33}$ erg s$^{-1}$. Larger flare contributions would be inconsistent with estimates that approximately 90\% of the quiescent emission from \sgra\ is in fact associated with spatially resolved emission.

\begin{deluxetable}{lcc}
\tablewidth{5in}
\tablecaption{\chandra/\nustar Joint Spectral Parameters\label{tab:xrayflares}}
\tablehead{\colhead{Parameter} & \colhead{Value} & \colhead{Units}}
\startdata
$N_{\rm H}$ & $17.8_{-2.5}^{+3.5}$ & $10^{22}$\,cm$^{-2}$\\
 $\Gamma$ & $2.1_{-0.4}^{+0.5}$ & \dots \\
 $N_{\rm FPMB}$ & $1.2_{-0.4}^{+0.9}$ & \dots \\
 $K_{\rm vapec}$ & $0.0012_{-0.0005}^{+0.0009}$ & \dots \\
 $kT_{\rm vapec}$ & $2_{-0.5}^{+0.4}$ & keV\\
 $N_{\rm ACIS}$ & $1_{-0.3}^{+0.7}$ & \dots \\
 $F_{\rm 0704,2-10}$ & $0.3_{-0.2}^{+0.2}$ & $10^{-12}$\,erg\,s$^{-1}$\,cm$^{-2}$\\
 $F_{\rm 0704,3-79}$ & $0.5_{-0.4}^{+0.8}$ & $10^{-12}$\,erg\,s$^{-1}$\,cm$^{-2}$\\
  $F_{\rm 1104,2-10}$ & $7.8_{-3.9}^{+2.6}$ & $10^{-12}$\,erg\,s$^{-1}$\,cm$^{-2}$\\
$F_{\rm 1104,3-79}$ & $15.4_{-7.5}^{+8.9}$ & $10^{-12}$\,erg\,s$^{-1}$\,cm$^{-2}$\\
 $F_{\rm Q,2-10}$ & $0.5_{-0.1}^{+0.2}$ & $10^{-12}$\,erg\,s$^{-1}$\,cm$^{-2}$\\
$F_{\rm Q,3-79}$ & $0.31_{-0.03}^{+0.05}$ & $10^{-12}$\,erg\,s$^{-1}$\,cm$^{-2}$\\
 \enddata
\tablecomments{$N_{\rm H}$ is the X-ray absorbing column density. $\Gamma$ is the flare photon index. $N_{\rm FPMB}$ is the cross-normalization of the \nustar FPMB relative to the FPMA. $K_{\rm vapec}$ and $kT_{\rm vapec}$ are the normalization and temperature of the quiescent {\tt vapec} component. $N_{\rm ACIS}$ is the cross-normalization of the \chandra ACIS spectrum relative to \nustar. Flare ($F_{0704},$ $F_{1104}$) and quiescent fluxes ($F_{\rm Q}$) are quoted for the intervals 2--10\,keV and 3--79\,keV --- the X-ray quiescent and flare fluxes presented in 
Table \ref{tab:sed_2017} are $\nu$F$_{\nu}$ in units of erg s$^{-1}$ cm$^{-2}$, and so differ slightly from the integrated values quoted here. See \S \ref{sec:joint-xray} and \S \ref{subs:sgra_sed} for further details.}
\end{deluxetable}

\section{Final Data Products}
\label{sec:data}

\subsection{Event Horizon Telescope Data Products} 
\label{subs:eht_data}

In this section, we describe the properties of the 2017 \sgra \ac{eht} \ac{vlbi} data. In particular, we detail the $(u,v)$-coverage, correlated flux densities, systematic error budgets, estimations for residual antenna-based gain errors, the influence of interstellar scattering on the measured visibilities, conservative estimates for the size of \sgra at 1.3\,mm, 
and an assessment of the source variability. 

\subsubsection{Data content}

\begin{figure*}
\centering
\includegraphics[width=0.85\textwidth]{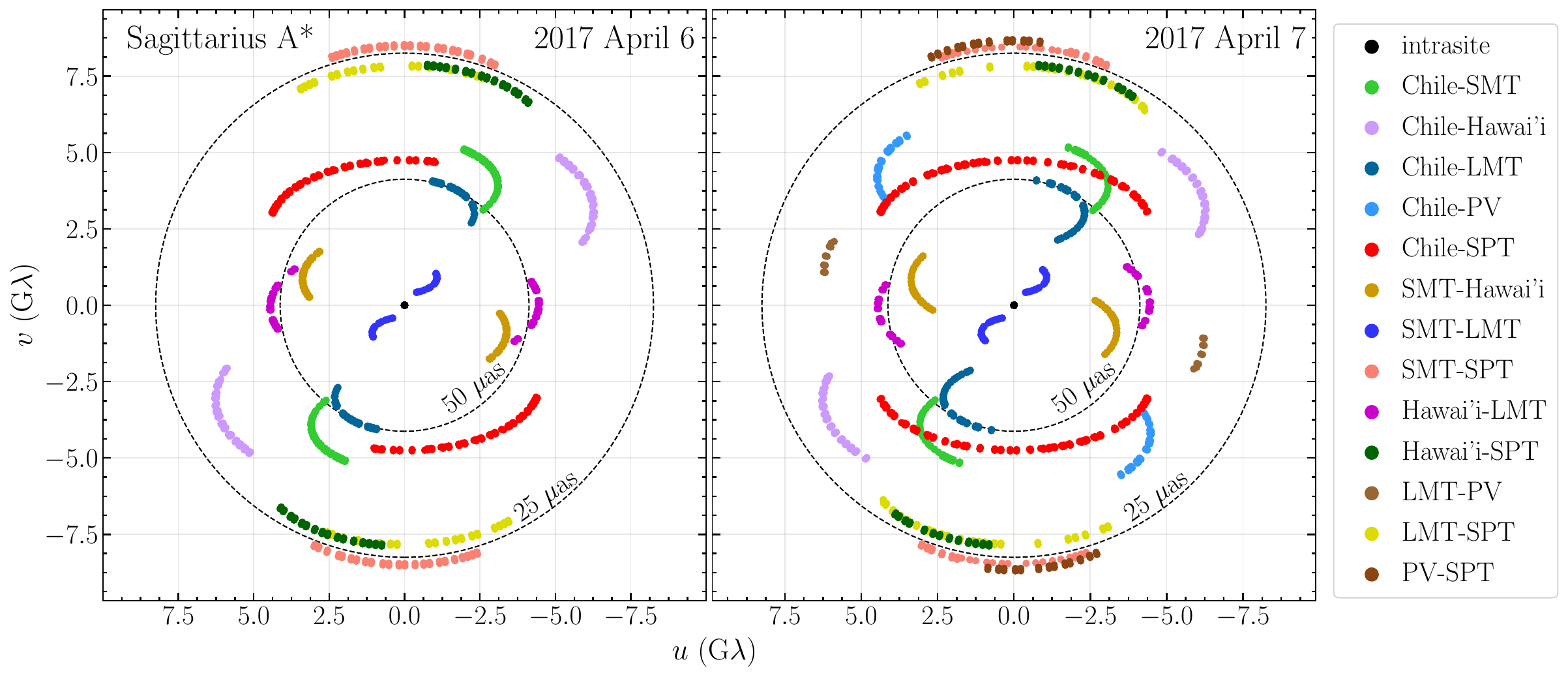}
\caption{$(u, v)$-coverages from fringe detections of \sgra on 2017 April 6 and 7 plotted from scan-averaged data. Fringe spacings of 25 and 50\,\ac{muas} are indicated with dashed circles.
}
\label{fig:detections}
\end{figure*}

\begin{figure*}
\centering
\includegraphics[width=0.9\textwidth]{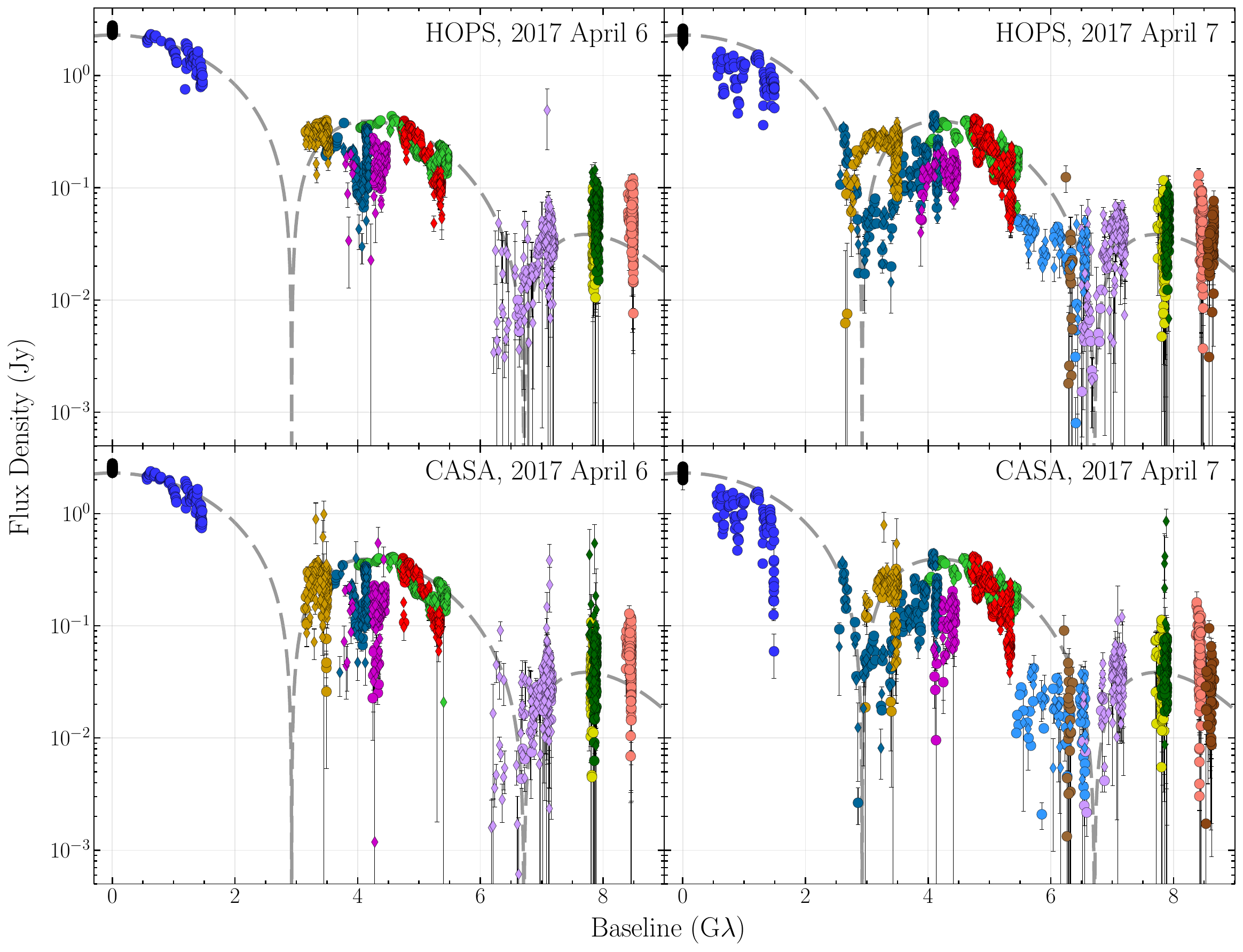}
\caption{Measured correlated flux densities of \sgra on April 6 and 7 of the HOPS (top panel) and CASA (bottom panel) data overplotted with a thin (unresolved) ring model with a \sdiam 
diameter, that has been blurred with a 23\,\ac{muas} FWHM circular Gaussian source \citepalias{PaperIII, PaperIV}.
Flux densities have been calibrated based on estimated station \ac{sefd}s, measured light curves (Section~\ref{sub:minispiral_cal}), and calibrator gain transfers (Section~\ref{sec:gainerrors}).
The data are averaged in 120\,s bins, which causes phase coherence losses in a few scans. Error bars denote $\pm 1\,\sigma$ uncertainty from thermal noise. Detections are color-coded by baselines, as in \autoref{fig:detections}. Redundant baselines are shown with different symbols: circles for baselines to \ac{aa} and \ac{sm}; diamonds for baselines to \ac{ap} and \ac{jc}.
}
\label{fig:eht-fluxes}
\end{figure*}

The observing schedules and $(u, v)$-coverages of our April 6 and 7 \sgra observing days are shown in \autoref{fig:schedules} and \autoref{fig:detections}, respectively.
Correlated flux densities as a function of baseline length are shown in \autoref{fig:eht-fluxes}. The measurements show a good agreement between the products of our processing pipelines. 
The amplitude data presented in \autoref{fig:eht-fluxes} are broadly consistent 
with the Fourier signature of a blurred ring, indicating the presence of two minima at $\sim$3\,G$\lambda$ and $\sim$6.5\,G$\lambda$.

The calibrated Stokes $\mathcal{I}$ VLBI data of \sgra are made publicly available through the EHT data portal\footnote{\url{https://eventhorizontelescope.org/for-astronomers/data}.} under the 2022-DXX-XX code.



\subsubsection{Systematic error budget}

\begin{table*}
\caption{Non-closing systematic uncertainties, $s$ (and in units of thermal noise, $s/\sigma_{\rm th}$), for \sgra and its calibrators estimated using various statistical tests on both the CASA and HOPS products. The data from April 6 and 7 have been used for \sgra, and for the calibrators, we have also used April 11. $n$ is the number of closure quantities used for each test.}\label{tab:syserr}
\begin{center}
\begin{tabularx}{\linewidth}{@{\extracolsep{\fill}}llcccccc}
\hline 
\hline 
 Source & Test & \multicolumn{3}{c}{CASA} & \multicolumn{3}{c}{HOPS}\\
 \cline{3-5}\cline{6-8}
 & & $s$ & $s/\sigma_{\rm th}$ & $n$ & $s$ & $s/\sigma_{\rm th}$ & $n$ \\
\hline
\hline
\sgra& RR\,$-$\,LL closure phases &  3.6$\degr$ & 0.6 & 313 &  3.6$\degr$ & 0.6 & 328 \\
& lo\,$-$\,hi closure phases &  2.1$\degr$ & 0.3 & 333 &  2.0$\degr$ & 0.3 & 367 \\
& trivial closure phases &  0.2$\degr$ & 0.1 & 373 &  1.0$\degr$ & 0.2 & 411 \\
& RR\,$-$\,LL log closure amplitudes & 10.9\% & 0.8 & 239 & 11.3\% & 0.9 & 278 \\
& lo\,$-$\,hi log closure amplitudes &  9.7\% & 0.6 & 284 &  7.6\% & 0.5 & 371 \\
& trivial log closure amplitudes &  5.6\% & 0.5 & 168 &  1.4\% & 0.1 & 212 \\
& lo $-$ hi closure trace phases &  1.1$\degr$ & 0.2 & 97 &  0.0$\degr$ & 0.0 & 125 \\
& trivial closure trace phases &  1.5$\degr$ & 0.3 & 160 &  0.0$\degr$ & 0.0 & 158 \\
& lo $-$ hi log closure trace amplitudes &  3.7\% & 0.4 & 97 &  0.0\% & 0.0 & 125 \\
& trivial log closure trace amplitudes &  5.7\% & 0.6 & 160 &  4.6\% & 0.5 & 158 \\
\hline 
NRAO\,530& RR\,$-$\,LL closure phases &  1.4$\degr$ & 0.2 & 125 &  0.0$\degr$ & 0.0 & 121 \\
& lo\,$-$\,hi closure phases &  2.5$\degr$ & 0.3 & 156 &  2.8$\degr$ & 0.3 & 150 \\
& trivial closure phases &  0.0$\degr$ & 0.0 & 151 &  0.0$\degr$ & 0.0 & 147 \\
& RR\,$-$\,LL log closure amplitudes &  0.4\% & 0.0 & 114 &  0.0\% & 0.0 & 117 \\
& lo\,$-$\,hi log closure amplitudes &  0.0\% & 0.0 & 203 &  2.7\% & 0.1 & 188 \\
& trivial log closure amplitudes &  6.2\% & 0.6 & 100 &  3.2\% & 0.3 & 91 \\
& lo $-$ hi closure trace phases &  1.3$\degr$ & 0.2 & 56 &  0.0$\degr$ & 0.0 & 54 \\
& trivial closure trace phases &  0.0$\degr$ & 0.0 & 104 &  0.0$\degr$ & 0.0 & 95 \\
& lo $-$ hi log closure trace amplitudes &  5.6\% & 0.5 & 56 &  8.5\% & 0.8 & 54 \\
& trivial log closure trace amplitudes &  2.1\% & 0.2 & 104 &  0.0\% & 0.0 & 95 \\
\hline 
J1924$-$2914& RR\,$-$\,LL closure phases &  2.4$\degr$ & 0.5 & 207 &  2.6$\degr$ & 0.5 & 235 \\
& lo\,$-$\,hi closure phases &  1.1$\degr$ & 0.2 & 224 &  0.9$\degr$ & 0.2 & 280 \\
& trivial closure phases &  0.5$\degr$ & 0.2 & 190 &  0.0$\degr$ & 0.0 & 204 \\
& RR\,$-$\,LL log closure amplitudes &  2.0\% & 0.2 & 272 &  3.5\% & 0.3 & 313 \\
& lo\,$-$\,hi log closure amplitudes &  6.6\% & 0.6 & 367 &  6.8\% & 0.7 & 455 \\
& trivial log closure amplitudes &  3.7\% & 0.5 & 190 &  4.2\% & 0.6 & 218 \\
& lo $-$ hi closure trace phases &  2.3$\degr$ & 0.5 & 222 &  1.3$\degr$ & 0.3 & 253 \\
& trivial closure trace phases &  2.2$\degr$ & 0.8 & 106 &  1.9$\degr$ & 0.7 & 106 \\
& lo $-$ hi log closure trace amplitudes &  5.5\% & 0.7 & 222 &  6.4\% & 0.8 & 253 \\
& trivial log closure trace amplitudes &  0.0\% & 0.0 & 106 &  2.6\% & 0.5 & 106 \\
\hline
\hline
\end{tabularx}
\end{center}
\end{table*}

\begin{figure*}
\centering
\begin{tabular}{cc}
\includegraphics[width=0.5\textwidth]{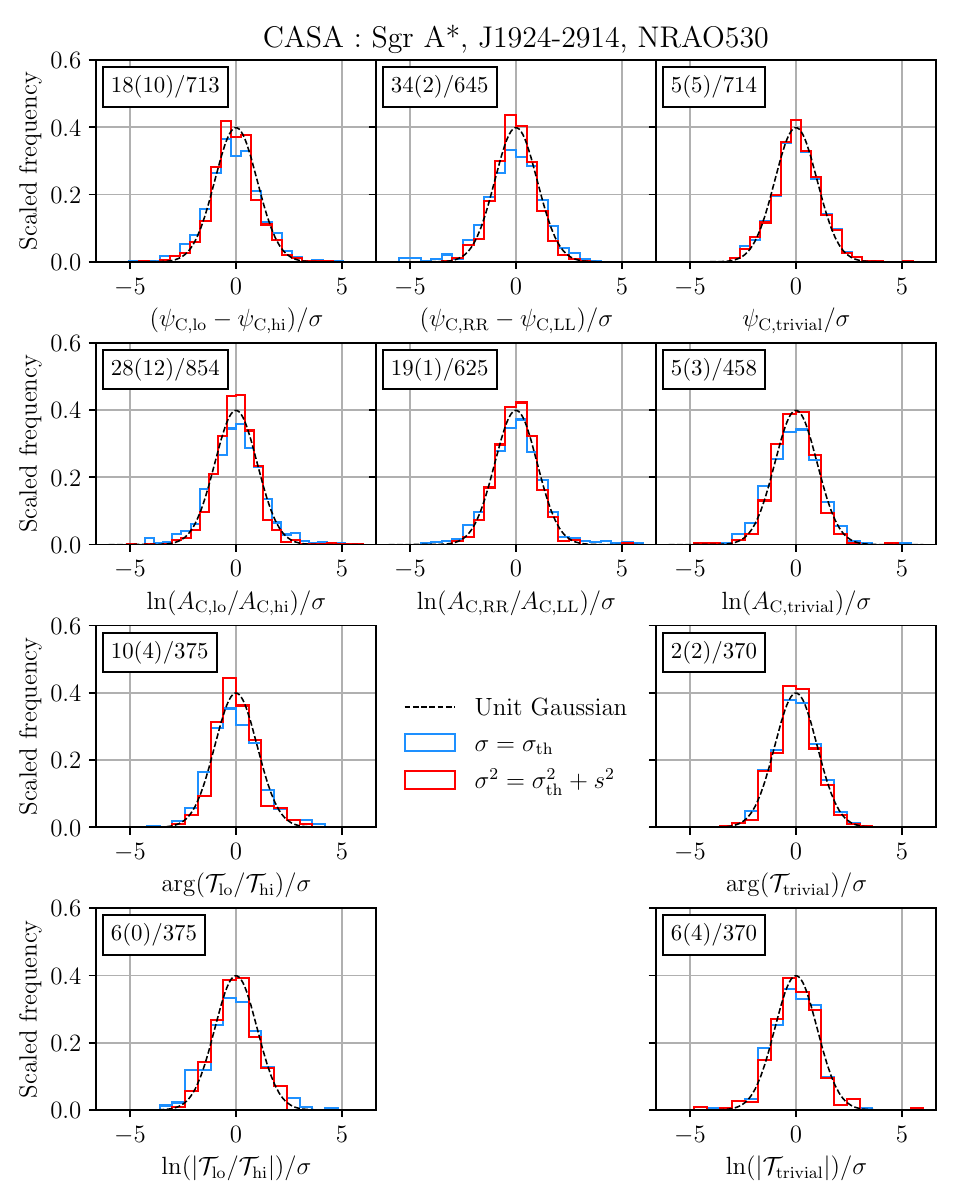} &
\includegraphics[width=0.5\textwidth]{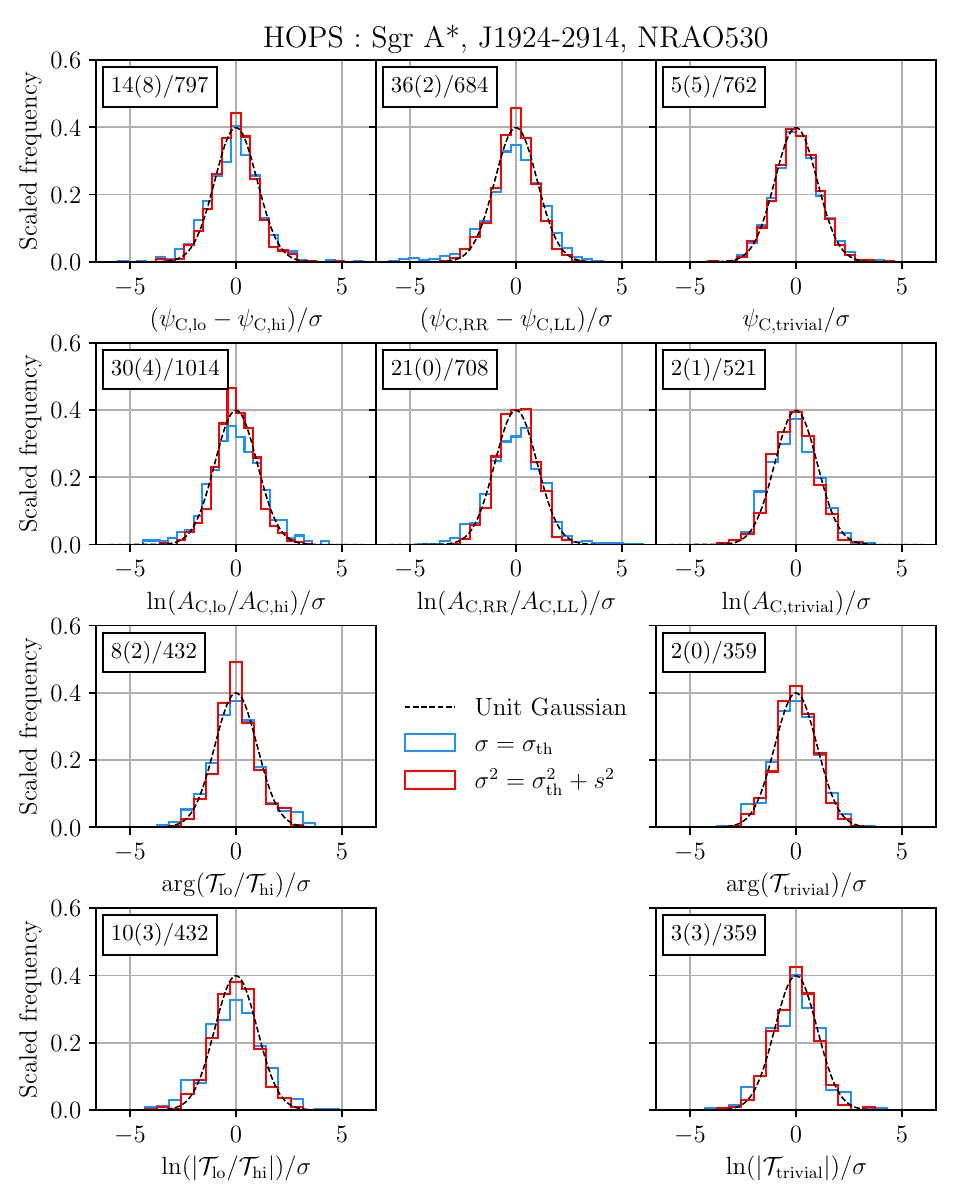}
\\(a)&(b)
\end{tabular}
\caption{Normalized distributions of lo$-$hi, RR$-$LL and trivial closure quantities, as well as lo$-$hi and trivial closure trace quantities of \sgra, NRAO\,530, and J1924$-$2914 from both the CASA (a) and HOPS (b) reduction pipelines. The data from April 6 and 7 have been used for \sgra, and for the calibrators, we have also used April 11. The distributions prior to (blue) and after (red) accounting for the estimated systematic uncertainties, $s$, are shown. The values of $s$ for each source and reduction pipeline are given in Table~\ref{tab:syserr}. In the top left corner of each distribution, the number of $>3\,\sigma$ outliers are given considering thermal noise only followed by the number of outliers considering thermal plus systematic noise for $\sigma$ in parenthesis. These numbers are followed by the total number of data points after a slash.
}
\label{fig:syserr}
\end{figure*}

We compute \ac{sn}\,$>$\,7 closure phases ($\psi_C$) and log closure amplitudes ($\ln{A_C}$), which are reasonably well described by a normal distribution \citep{2020Blackburn}, to estimate the amount of systematic noise $s$ present in the \ac{eht} data following \citetalias{M87PaperIII} and \citet{Wielgus2019}.
We have augmented these ``traditional'' closure phases and closure amplitude tests with novel closure trace quantities, $\mathcal{T}$, described in \citet{ClosureTraces2020}. Closure traces are complex composite data structures, which we characterize by their phase and log amplitude. The closure traces are produced from parallel- and cross-hand correlation products, and they are insensitive to all linear station-based corruptions of the data, including both station gains and polarization leakage.
The uncertainty $\sigma_x$ of a data product $X$ follows from the a priori estimated thermal noise $\sigma_\mathrm{th}$ from our calibration pipelines and a constant systematic non-closing error $s_x$ as
\begin{equation}
    \sigma^2_x = \sigma_\mathrm{th}^2 + s_x^2 \,.
\end{equation}
We estimate $s_x$ based on the criterion that the median absolute deviation of $X/\sigma_x$ becomes unity for the following data quantities:
\begin{enumerate}
    \item RR\,$-$\,LL closure phases ($X = \psi_\mathrm{C,\,RR} - \psi_\mathrm{C,\,LL}$) and log closure amplitudes ($X = \ln{A_\mathrm{C,\,RR}} - \ln{A_\mathrm{C,\,LL}}$), which should be zero in the absence of significant circular source polarization and instrumental polarization leakage.
    \item Low-band\,$-$\,high-band closure phases ($X = \psi_\mathrm{C,\,lo} - \psi_\mathrm{C,\,hi}$, log closure amplitudes ($X = \ln{A_\mathrm{C,\,lo}} - \ln{A_\mathrm{C,\,hi}}$), closure trace phases ($X = {\rm arg}({\mathcal{T}_{\rm lo}})-{\rm arg}({\mathcal{T}_{\rm hi}})$), and log closure trace amplitudes ($X = \ln{|\mathcal{T}_{\rm lo}|}-\ln{|\mathcal{T}_{\rm hi}|}$), all of
    which should be close to zero in the absence of significant variations of source structure and flux over the observed \ac{eht} bandwidth. The phase shift due to the small frequency difference between the two bands is negligible.
    \item Closure phases of small, trivial triangles ($X = \psi_\mathrm{C,\,trivial}$) and log closure amplitudes of small, trivial quadrangles ($X = \ln{A_\mathrm{C,\,trivial}}$), which should be zero in the absence of significant instrumental polarization leakage as they probe the symmetric large-scale source structure.
    Additionally, trivial closure phases will be affected by large-scale source structures.
    \item Closure trace phases ($X = {\rm arg}({\mathcal{T}_{\rm trivial}})$) and closure trace log amplitudes ($X = \ln{|\mathcal{T}_{\rm trivial}|}$) on trivial boomerang quadrangles, in which a site is repeated so that the area of the quadrangle vanishes.  Both sets of quantities should be zero in the absence of nonlinear station-based or baseline-based errors.
\end{enumerate}

To avoid biases in the closure trace log amplitudes arising from low \ac{sn}, we construct all closure quantities $X$ from visibilities that have been coherently averaged over 120\,s and apply the same \ac{sn}\,$>$\,7 threshold; we subsequently average the constructed closure traces on scan-length intervals.
We note that these averaging timescales carry with them the potential for decoherence losses resulting from atmospheric delays and/or structural variability.

The systematic error budgets derived for \sgra, NRAO\,530, and J1924$-$2914 are shown in \autoref{tab:syserr}. These are estimated using only data on days when \ac{aa} was observing, i.e., 2017 April 6, 7, and 11. However, we exclude \sgra data from 2017 April 11 due to flaring activity -- these will be the subject of a forthcoming study \citep[see also][]{Wielgus2022}. The underlying distributions of the various data quantities, with and without added systematic error, are shown in \autoref{fig:syserr}.

From the combined data on all baselines, an excess in systematic errors of RR\,$-$\,LL closure quantities can be seen for \sgra compared to the observed NRAO\,530 and J1924$-$291 calibrator sources.
These offsets may be indicative of intrinsic source polarization in \sgra. However, the overall offsets are just at the level of thermal noise for scan-averaged data.  
It is likely for source polarization to be significant on particular baselines only. Horizon-scale polarization signatures in the 2017 \sgra \ac{eht} data will be analyzed in future work.

The systematic error budgets agree between the traditional closure quantities and the novel closure traces.
It is therefore unlikely for large-scale source structure and uncorrected polarization leakage effects to be the dominant sources of systematic uncertainties.

There are several \ac{eht} data issues affecting \sgra observations that require special care.
These are described in detail in \citetalias{M87PaperIII} and briefly summarized below.
\begin{enumerate}
\item The \ac{jc} and \ac{sm} used an identical frequency setup derived from a shared frequency standard in 2017. Radio-frequency interference (RFI) is not washed out due to low fringe rate. Both stations are therefore never chosen as the reference station for the correction of atmospheric phase fluctuations.
Resultant amplitude errors on this baseline are mostly mitigated by flagging channels affected by RFI and network calibration.

\item An instability in the maser used in 2017 for \ac{pv} caused a de-correlation over the 0.4\,s correlator accumulation period, which was corrected by up-scaling the \ac{sefd}.

\item Partial data dropouts due to a misconfigured Mark 6 recorder at \ac{ap} have been accounted for by adjusting the amplitudes and data weights accordingly during correlation. A hard drive failure at the \ac{jc}, causing 1/16th of the low band data to be lost, has been similarly corrected.
Furthermore, a small \ac{sefd} correction factor has been applied to the \ac{ap} data to correct for an amplitude loss from an interfering 1 pulse-per-second signal.

\item High band \ac{sm} \ac{sefd}s on the first three observing days have been up-scaled to correct for occasionally corrupted frequency channels in the beamformer system.

\item A negligible $< 0.1$\,\% amplitude loss due to a periodic \ac{aa} correlator glitch occurring every 18.192\,s was left uncorrected.
Moreover, a baseline-dependent $\sim 0.67$\,\% signal loss due to finite fast-Fourier-transform lengths over the full 2048\,MHz band on non-\ac{aa} baselines has not been corrected.

The systematic error studies presented in \autoref{fig:syserr} and \autoref{tab:syserr} reflect a comparable level of data quality and internal self-consistency for the CASA and HOPS data products. All data distributions are well behaved and the different amounts of estimated systematic noise are balanced, being sometimes higher in one or the other data set.
Small differences in the number of recorded visibilities are due to the effects of thresholding near the detection limit as well as slight differences in flagging between the two pipelines.

As indicated also by the scatter in the $s$ measurements, optimal systematic error budgets depend on the exact baselines and data quantities used. A crude recommendation for data averaged in 120\,s bins is to adopt \ang{2.5} for closure phases and 9\,\% for log closure amplitudes.

\end{enumerate}

\subsubsection{Gain errors}
\label{sec:gainerrors}

\begin{table*}
\caption{\ac{eht} flux density calibration parameters and their uncertainties. For phased arrays (ALMA and SMA), the DPFUs represent
the combined sensitivity of all phased dishes.
The gain curve parameters as a function of elevation $E$ are given based on a $\mathrm{gc}(E) = 1 - B (E - E_0)^2$ parameterization (\autoref{eq:gcfit}).
}             
\label{tab:ER6uncertainties}      
\begin{tabularx}{\linewidth}{@{\extracolsep{\fill}}l c c c c}
\hline\hline       
                      
Station(code) & RCP DPFU [K/Jy] & LCP DPFU [K/Jy] & $B$ & $E_0$\\ 
\hline  
   ALMA(AA)\textsuperscript{a} & 1.03 $\pm$ 10\,\% & 1.03 $\pm$ 10\,\% & 0 & 0\\
   APEX(AP) & 0.0245 $\pm$ 11\,\% & 0.0250 $\pm$ 11\,\% & 0.00002 $\pm$ 3.6\,\% & 36.6 $\pm$ 1\,\%\\
   IRAM~30-m(PV)\textsuperscript{b} & 0.034 $\pm$ 10\,\% & 0.033 $\pm$ 10\,\% & 0.00018 $\pm$ 5.3\,\% & 43.7 $\pm$ 1.3\,\%\\ 
   JCMT(JC)\textsuperscript{c} & (0.026 $\pm$ 14\,\%) -- (0.033 $\pm$ 11\,\%) & (0.026 $\pm$ 14\,\%) -- (0.033 $\pm$ 11\,\%) & 0 & 0\\
   LMT(LM)\textsuperscript{d} & 0.061 $\pm$ 35\,\% & 0.064 $\pm$ 35\,\% & 0 & 0\\
   SMA(SM)\textsuperscript{e} & 0.046 $\pm$ (5 -- 15)\,\% & 0.046 $\pm$ (5 -- 15)\,\% & 0 & 0\\
   SMT(AZ) & 0.01683 $\pm$ 7\,\% & 0.01681 $\pm$ 7\,\% & 0.000082 $\pm$ 10.4\,\% & 57.6 $\pm$ 2.0 \,\%\\
   SPT(SP)\textsuperscript{f} & 0.0061 $\pm$ 15\,\% & 0.0061 $\pm$ 15\,\% & 0 & 0\\
\hline
\end{tabularx}
\newline
\textsuperscript{a}\footnotesize{The ALMA DPFU uncertainty is based on the overall 10\,\% uncertainty estimated by the QA2 team.}\newline
\textsuperscript{b}\footnotesize{The PV DPFU has been scaled down by a factor of 3.663 to account for the known maser instability in 2017.}\newline
\textsuperscript{c}\footnotesize{For the JCMT DPFU, a range between the smallest daytime value and the nighttime value is given.}\newline
\textsuperscript{d}\footnotesize{The LMT has an unparametrized 10\,\% uncertainty on the gain curve, which has been added to the DPFU uncertainty. The DPFU values shown here are from 2017 April 7. The uncertainty is the most conservative value from April 6.}\newline
\textsuperscript{e}\footnotesize{The SMA DPFU uncertainty is based on the dominant 5-15\,\% uncertainty on the phasing efficiency.}\newline
\textsuperscript{f}\footnotesize{The gain curve of the SPT is uncharacterized, as normalized antenna temperature values cannot be obtained from sources that remain at a constant elevation when observed from the South Pole.}
\end{table*}

Ensembles of opacity-corrected antenna temperature ($T_a^*$) measurements are used to fit for gain curves and DPFUs \citep{Janssen2019b, Issaoun2019}.
For the gain curve, we fit 
\begin{equation}
    \mathrm{gc}(E) = 1 - B \left( E - E_0 \right)^2,
    \label{eq:gcfit}
\end{equation}
to normalized $T_a^*$ values from quasars tracked over a wide range of elevations $E$. The parameters $B$ and $E_0$ describe the peak and shape of the gain curve, respectively.
For the DPFU, we typically fit a constant value for the entire observing track to $T_a^*$ measurements of solar system objects to estimate the constant aperture efficiency $\eta_\mathrm{ap}$.
A statistical $T_a^*$ scatter translates into uncertainties of the fitted $B$, $E_0$, and $\eta_\mathrm{ap}$ parameters of the gain curve and DPFU, respectively.
The uncertainty of the planet model brightness temperature is added in quadrature for the DPFU fitting.
For the \citet{Butler-planet-models} models, we have a 10\,\% uncertainty for Saturn and 5\,\% for Mars, Jupiter, and Uranus.
The error contribution from $T_\mathrm{sys}^*$ measurements is negligible.
Finally, \ac{aa} and \ac{sm} exhibit an additional calibration uncertainty from the phasing efficiency.
The performance of \ac{aa} has been determined in the quality assurance stage 2 (QA2) ALMA interferometric reduction of data  \citep{2019PASP..131g5003G}. The final calibration method is described in Section~\ref{sub:minispiral_cal}, for which we have estimated a global 10\,\% flux density calibration uncertainty.
The small \ac{sm} dishes are well characterized. Here, the phasing efficiency dominates the overall calibration uncertainty, which ranges between 5\,\% in optimal observing conditions and 15\,\% when the phasing efficiency is low.
The a priori flux calibration parameters are summarized in \autoref{tab:ER6uncertainties}.
The reported uncertainties provide a good upper limit for residual gain errors. 
Compared to our 2017 \m87 data \citepalias{M87PaperIII}, the DPFU values are slightly updated, which does not affect our previous results, while the gain curves stayed the same.

However, there are several loss factors that are not captured in our a priori calibration framework, most notably imperfect pointing and focus solutions of individual telescopes. The case is most severe for the large \ac{lm} dish, where the measured DPFU varies from day to day. In order to provide a conservative estimate of the gain uncertainties, a bootstrap approach was applied to the DPFU measurements on 2017 April 6 and 7. With 10000 medians drawn from the bootstrapped samples, we are able to derive the Median Absolute Deviations, which is then scaled to an equivalent sigma.
With this method we find a $\sim35\,\%$ uncertainty on the \ac{lm} DPFU.

\begin{figure}
\includegraphics[width=\columnwidth]{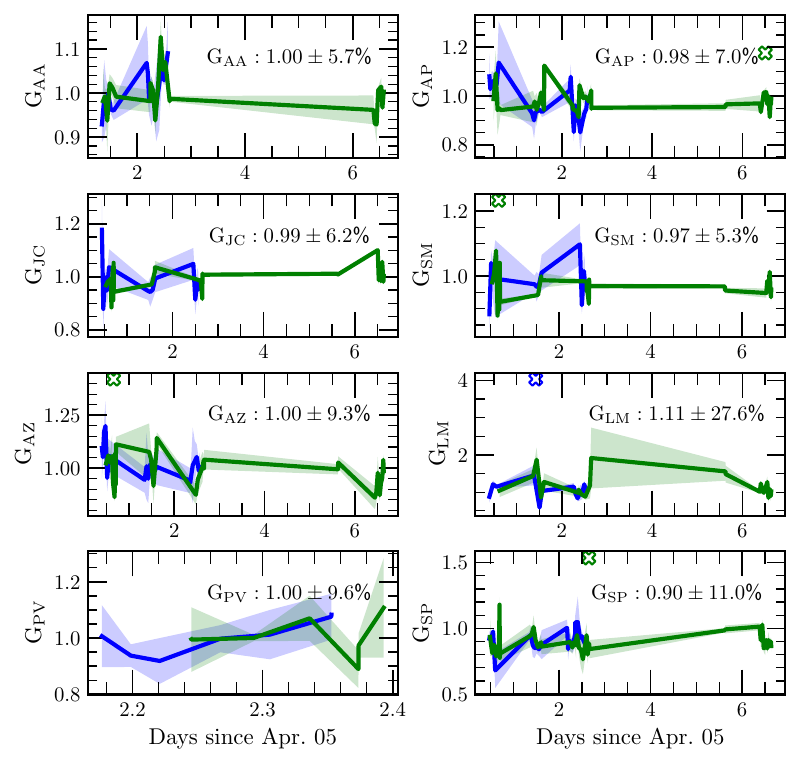}
\caption{Low-band antenna gains of NRAO\,530 (blue curve) and J1924$-$2914 (green curve) by amplitude self-calibration. The two-letter station codes are taken from \autoref{tab:ER6uncertainties}. The colored bands indicate $1\,\sigma$ uncertainties of the gains. The empty cross symbols represent the flagged data points lying beyond $3\,\sigma$ from the mean value. The PV telescope observed \sgra only on April 7.  
}
\label{fig:gains_cals}
\end{figure}

Additionally, the antenna gains can be characterized by amplitude self-calibration, which solves the empirical corrections for time-variable instrumental or environmental factors that cannot be measured directly. In order to avoid any variability from \sgra that could contaminate the antenna gain estimations, the scan-averaged visibility data of calibrators (J1924$-$2914 and NRAO\,530) are utilized by assuming their stationarity in both source structure and flux density. For both calibrators, the fiducial images independently produced with different imaging pipelines \citepalias[i.e., eht-imaging, SMILI, Difmap and DMC;][]{PaperIII, PaperIV} are employed to improve the statistics on the gains (for more details, see \citet{eht2017-J1924, eht2017-NRAO530}). We use the mean values of the gains obtained from each imaging pipeline. As an example, the resultant antenna gains derived by amplitude self-calibration for the low-band datasets for both calibrators are shown in \autoref{fig:gains_cals}. 
The results were obtained after performing a gain alignment procedure meant to minimize the offset between the gains from the two calibrators within the overlapping time and flagging the data points lying beyond $3\sigma$ from the mean value. The procedure is as follows: (1) we first derive a mean gain value for each calibrator by only using the gains in the overlapping times; (2) from these gain mean values for the two calibrators, in the overlapping times, we derive an average gain value and re-scale the two calibrators' mean gains with this average value; (3) the scaling factors thus obtained are then applied to the gains over the entire time range.
The mean gains for each station (reported on top of each frame in \autoref{fig:gains_cals}) are within their corresponding a priori DPFU error budgets (\autoref{tab:ER6uncertainties}). We can therefore assume that the calibrator self-calibration method captures the known gain uncertainties that enter into our a priori flux density calibration error budget. At this point, we linearly interpolate the fiducial gains obtained from the calibrators to the \sgra timestamps and we apply them to the \sgra data.


\subsubsection{Interstellar scattering}
\label{sec:scattering}


\begin{figure*}[t]
    \centering
    \begin{tabular}{cc}
        \includegraphics[height=6cm]{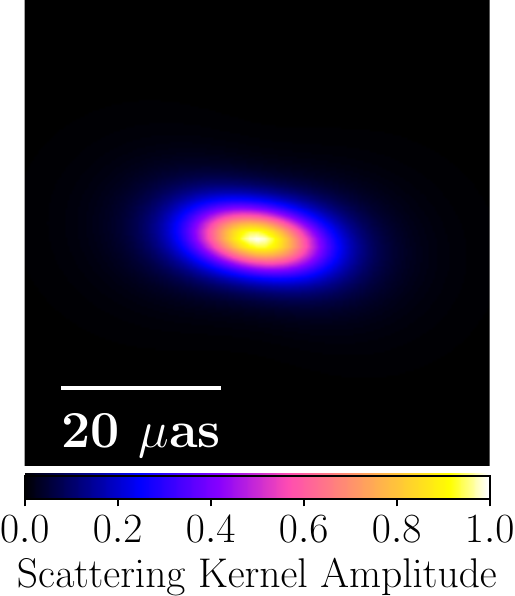} &
        \includegraphics[height=6cm]{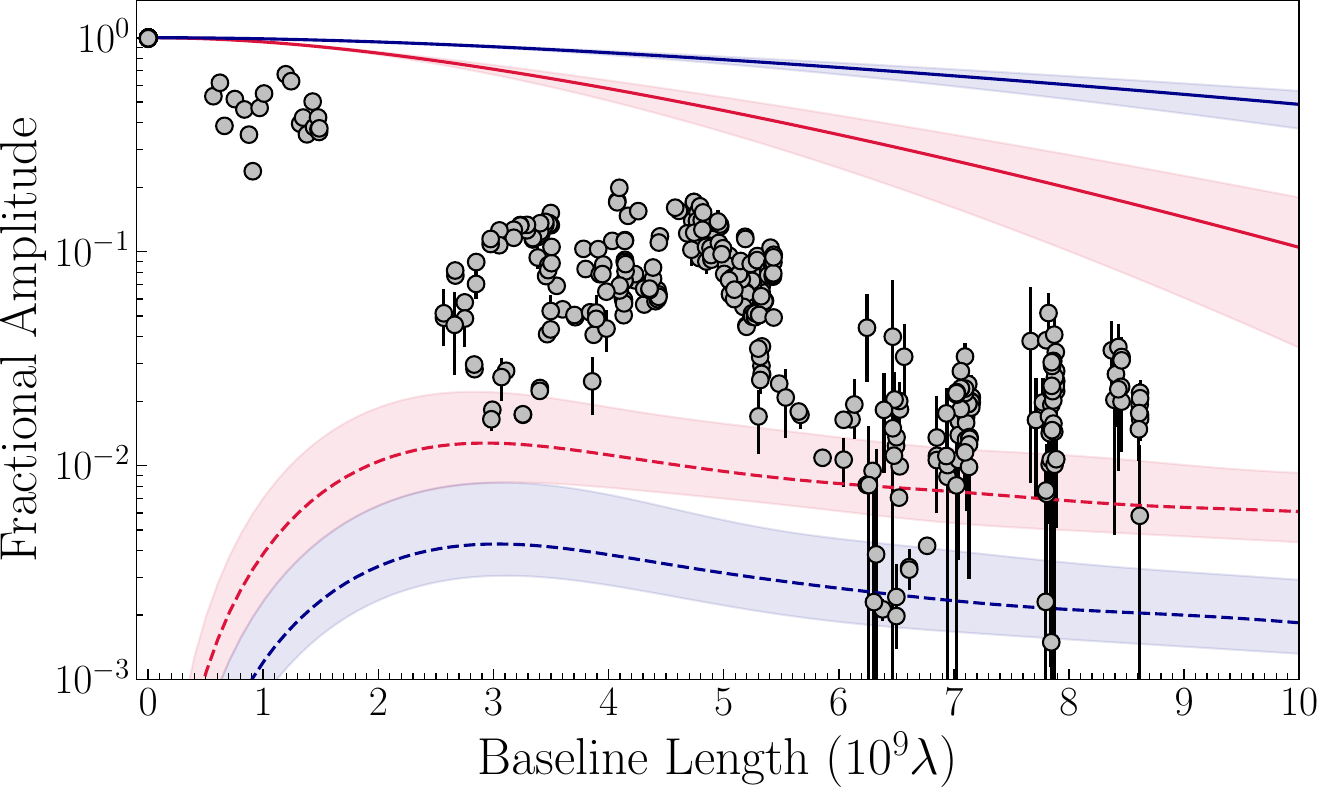}\\
        (a)&(b)
    \end{tabular}
    \caption{
    Diffractive scattering kernel and refractive noise rms at 1.3\,mm based on the scattering model of \citet{Psaltis_2018} and \citet{2018ApJ...865..104J}. (a) The image-domain representation of the kernel, with a second moment of $23.6 \times 12.1$\,${\rm \mu}$as in FWHM at PA=82$^\circ$. (b) The normalized kernel amplitude and refractive noise level at 1.3\,mm as a function of the baseline length (lines), overlaid on normalized \sgra low band data from 2017 April 7 (dots). 
    The red and blue solid lines show the amplitudes of the kernel along the major and minor axes, respectively.
    The dashed lines show the rms refractive noise levels for a reference model in \citetalias{PaperIII} along each axis colored in the same way.
    The shaded area around each line covers the range of possible amplitude values given the uncertainties of scattering parameters in the model.
    The \sgra amplitudes exhibit a steeper fall-off than that of the scatter-broadening kernel, indicative of resolved dominant intrinsic structure larger than the scattering kernel size.
    Furthermore, most of amplitudes are well above the expected range of the refractive noise, implying that the refractive substructure is likely not dominant in the observed structure.
    The most affected baselines are Chile to Hawai`i and Spain, Mexico to Spain and the South Pole, and Hawai`i to the South Pole in the 6\,--\,8\,G$\lambda$ range.
    }
    \label{fig:scatt_kernel}
\end{figure*}

Interstellar scattering is a long-known extrinsic effect on the observed radio structure of \sgra \citep{Davies_1976}. The scattering of radio waves from \sgra is predominantly caused by foreground stochastic turbulent electrons along the line of the sight located far from the Galactic Center \citep{Bower_2014,Dexter_2017}. 
At long observing wavelengths, the most dominant effect on the source structure is angular broadening, a chromatic effect arising from diffractive scattering \citep[see, e.g.,][]{Narayan_1992}, resulting in a Gaussian image with a size proportional to the squared wavelength for wavelengths $\lambda \gsim 1\,{\rm cm}$ \citep{Davies_1976,vanLangevelde_1992,Bower_2004,Shen_2005,Bower_2006,Johnson_2018,Lu_2018}. Furthermore, the scattering is anisotropic, with stronger angular broadening along the east-west direction compared to the north-south direction \citep{Frail_1994, Jauncey1989}. 
The angular broadening has a full width at half maximum (FWHM) of $(1.380 \pm 0.013)\lambda_{\rm cm}^2\,{\rm mas}$ along the major axis and $(0.703 \pm 0.013)\lambda_{\rm cm}^2\,{\rm mas}$ along the minor axis, with the major axis at a position angle $81.9^\circ \pm 0.2^\circ$ east of north \citep{2018ApJ...865..104J}. The intrinsic source angular size is also chromatic, with a $\theta_{\rm src} \propto \lambda$ dependence \citep{2018ApJ...865..104J}.
Consequently, observations at 1.3\,mm, where historical measurements of \sgra determined the size to be $\sim50-60\mu$as \citep{Doeleman08,Fish_2011,Lu_2018,2018ApJ...865..104J}, are in the regime where the intrinsic source structure is no longer sub-dominant to scattering and becomes the dominant structure in the image. 
Angular broadening can be described by a convolution of an unscattered image with a scattering kernel, or equivalently by a multiplication of the un-scattered, intrinsic interferometric visibilities by the appropriate Fourier-conjugate kernel \citep{GoodmanNarayan89}. 
More background information and reviews on interstellar scattering can be found in \citet{Rickett_1990}, \citet{Narayan_1992}, and \citet{TMS}.

Secondary effects arise from density irregularities in the interstellar medium, causing stochastic variations in the scattering, as well as diffraction effects.
These variations introduce substructure in the image that is not intrinsic to the source \citep{GoodmanNarayan89,Johnson_Gwinn_2015,Johnson_Narayan_2016,Psaltis_2018}. Scattering-induced substructure was first discovered in images of \sgra at 1.3\,cm by \citet{Gwinn_2014} and later observed at other wavelengths \citep{2018ApJ...865..104J,2019ApJ...871...30I,Issaoun_2021,Cho_2022}, giving additional constraints on the scattering properties of \sgra. This substructure is caused by modes in the scattering material on scales comparable to the image extent, so scattering models with identical scatter-broadening may still exhibit strong differences in their scattering substructure. The substructure manifests in the visibility domain as ``refractive noise'', which is an additive complex noise component with broad correlation structure across baselines and time \citep{Johnson_Narayan_2016}. Using observations of \sgra\ from 1.3\,mm to 30\,cm, \citet{2018ApJ...865..104J} have shown that the combined image broadening and substructure strongly constrain the power spectrum of density fluctuations, which is consistent with later observations that occurred close to the 2017 EHT observations of \sgra described in this work (Sections~\ref{subsubs:eavn_obs} and \ref{subsubs:gmva_obs}) and in \citet{2019ApJ...871...30I,Issaoun_2021} and \citet{Cho_2022}. 

Calibrated EHT data sets indicate that the intrinsic structure is dominant in measured visibilities, and that both diffractive and refractive scattering effects are limited, as anticipated from the empirically obtained scattering model and early EHT observations.
In \autoref{fig:scatt_kernel}, we show the scattering kernel in both the image and visibility domains based on the scattering parameters in \citet{2018ApJ...865..104J}, together with the calibrated EHT data. The analysis from \citet{2018ApJ...865..104J} implies a non-Gaussian kernel (solid lines) more compact than the conventional Gaussian kernel adopted in early literature.
Consequently, the angular broadening effect, i.e., multiplication of the intrinsic visibilities with the Fourier-conjugate kernel of scattering, causes a slight decrease in visibility amplitudes (and therefore the signal-to-noise ratio) by a factor of a few at maximum. 
The observed \sgra visibility amplitudes exhibit a steeper decrease with baseline length than the scattering kernel, suggesting that the intrinsic source structure, larger than the scattering kernel, is clearly resolved despite angular broadening. 

The refractive substructure is likely also not dominant in \sgra at 1.3\,mm. 
In \autoref{fig:scatt_kernel} (b), we show typical refractive substructures along with the major and minor axes of the anisotropic scattering kernel expected for a reference model, adopted in  \citetalias{PaperIII}, computed with \texttt{eht-imaging} based on the \citet{Psaltis_2018} and \citet{2018ApJ...865..104J} scattering model (see \citetalias{PaperIII} Section 4.1 for details).
While the rms noise levels due to the refractive substructure are dependent on the source size and shape \citep[see, e.g.,][]{Johnson_Narayan_2016}, the variations due to the deviation of the intrinsic source structure from the reference model are expected to be within a factor of a few \citepalias{PaperIII}. Consequently, the effects of the refractive structure remains only a few percent of the total flux density at the maximum regardless of the source structure, hardly affecting the measured visibilities except for data points around the second null and/or at baselines longer than $\sim 6$\,G$\lambda$, which are in the low \ac{sn} regime. A more detailed analysis of the scattering properties on the observed visibilities and source structure is provided in \citetalias{PaperIII}. For the purposes of the current paper, we conclude that an estimate of the size of \sgra drawn from visibility data will be dominated by the intrinsic structure of the source rather than scattering.


\subsubsection{Source size and fractional compact flux density}
\label{sec:source_size}

\begin{table*}[t!]
\caption{Constraints on fractional compact flux density and source size measured along the directions of the LMT--SMT and LMT--ALMA baselines.}
\label{tab:source_size}
\begin{center}
\begin{tabularx}{\linewidth}{@{\extracolsep{\fill}}ccccc}
\hline 
\hline 
UTC day      & Fractional Compact & Fractional Compact Flux Density & Minimum Source Size & Maximum Source Size \\
(2017 April) & Flux Density & (with the minimum source size) & ($\rm{\mu as}$) & ($\rm{\mu as}$)     \\
\hline 
$6$ & $0.83$ & $0.88$ & $42$ & $79$ \\
$7$ & $0.66$ & $0.73$ & $38$ & $95$ \\
$6+7$ & $0.83$ & $0.88$ & $39$ & $87$ \\
\hline
\hline
\end{tabularx}
\end{center}
\end{table*}

\begin{figure*}[t!]
\centering
\begin{tabularx}{\textwidth}{CC}
\includegraphics[height=6cm]{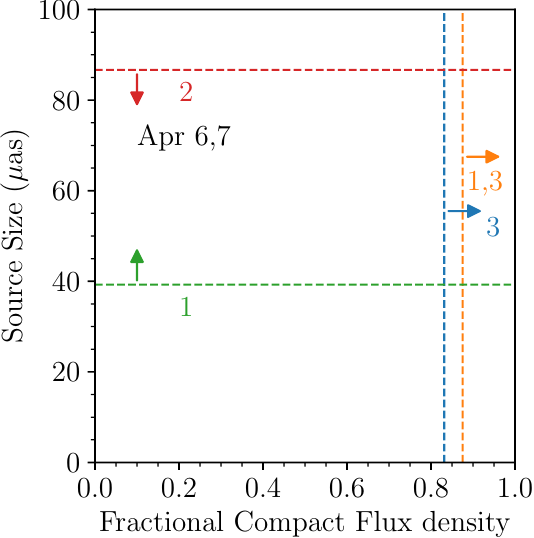} &
\includegraphics[height=6cm]{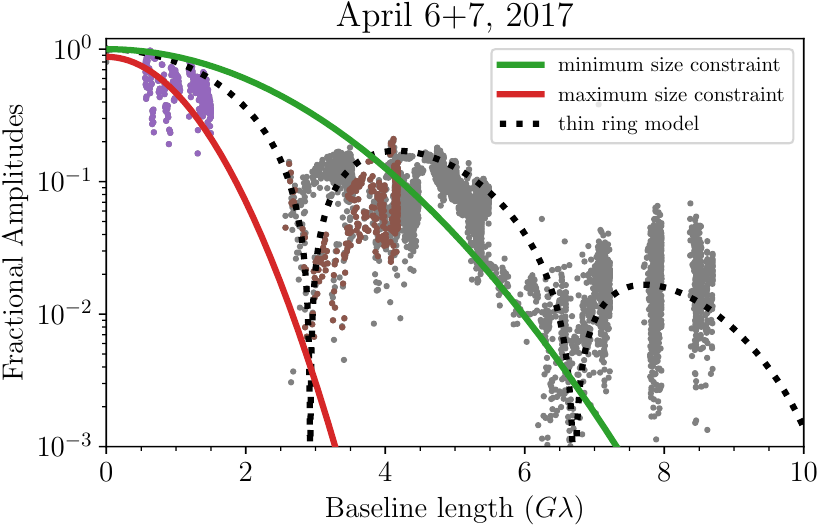} \\
(a) & (b)
\end{tabularx}
\caption{
Joint constraints on the size and total compact flux density of the 1.3\,mm emission in \sgra shown in \autoref{tab:source_size}. See \S\ref{sec:source_size}
for details.
(a) Numbered constraints correspond to: (1) the LMT--SMT/LMT--ALMA amplitude ratio, (2) LMT--SMT amplitudes coupled with the requirement $\mathrm{F}_{\rm cpt} \leq \mathrm{F}_{\rm tot}$, and (3) the maximum LMT--SMT amplitude. 
The position angles of the LMT--SMT and LMT--ALMA baselines range from $125^{\circ}$ to $164^{\circ}$.
(b)  
Fractional visibility amplitudes of circular Gaussian sources with FWHMs corresponding to the minimum (green line) and maximum (red line) size constraints for the April 6 and 7 data sets. Flux-normalized calibrated amplitudes are overlaid, where those of the LMT--SMT and ALMA--LMT baselines used to derive these size constraints are colored in purple and brown, respectively. The dotted line denotes the thin-ring model shown in \autoref{fig:eht-fluxes} for comparison.
}
\label{fig:source_size}
\end{figure*}

The 1.3\,mm source structure of \sgra has been reported to have FWHMs of several tens of microarcseconds from early EHT observations in the last decade \citep{Doeleman08,Fish_2011,Johnson_2015,Lu_2018}, consistent with our 2017 observations (\autoref{fig:detections}).
The SMT--LMT baseline is the shortest inter-site VLBI baseline provided in EHT observations, covering $\sim$0.6--1.5\,G$\lambda$ corresponding to a fringe spacing of $\sim$140--370\,\uas.
For a fringe spacing much larger than the source size, the visibility amplitude is well approximated by a quadratic function solely governed by the size along the baseline direction and the compact total flux density of the source \citep{Issaoun_2019b}. 
Along with the direction of the SMT--LMT baseline, the EHT array has the ALMA--LMT baseline with intermediate baseline lengths of 2.6-4.2\,G$\lambda$ (see \autoref{fig:detections}).
The visibility amplitudes on these baselines are useful to constrain the compact total flux density and source size on VLBI scales along the directions of these baselines together with the intra-site short baselines tracing the total flux density on arcsecond scales \citepalias[see][Appendix B.1]{M87PaperIV}.

Here, we derive constraints on the total compact flux density and the source size in a similar manner to \citetalias{M87PaperIV} by making use of these baselines. Compared with \m87 data presented in \citetalias{M87PaperIII}, \sgra data are further calibrated using interpolated gain solutions from calibrators \S\ref{sec:gainerrors}. 
We therefore modify the equations in \citetalias{M87PaperIV} and derive the three constraints outlined below.

{\it Constraint 1.}
The first constraint is based on the fact that the visibility amplitudes fall approximately quadratically with baseline length on short baselines, \citep{Issaoun_2019b}. 
The intermediate-to-long baselines tend to measure larger correlated flux density than what is expected for a Gaussian source with equivalent quadratic amplitude behavior on short baselines. While localized visibility-domain features such as ``nulls'' may give lower flux densities than the equivalent Gaussian, the presence of image substructure will tend to increase the average flux density.
On short baselines, we can thus express the visibility amplitudes in terms of an equivalent circular Gaussian visibility function,
\begin{align}
V_{\rm G}(\mathbf{u};I_0,\theta) = I_0 e^{- \frac{\left(\pi \theta |\mathbf{u}| \right)^2}{4\ln2}},
\end{align}
where $I_0$ is the total flux density of the Gaussian source, $|\mathbf{u}|$ is the length of the baseline, and $\theta$ is its FWHM in radians.
We can deduce that the measured amplitude ratio of the ALMA--LMT over SMT--LMT baslines will be larger than the corresponding ratio from a circular Gaussian source model.
Consequently, the FWHM size of a circular Gaussian determined by the amplitude ratio between SMT--LMT and ALMA--LMT baselines provides an estimate of the minimum compact source size $\theta_{\rm{cpt}}$ that is not significantly affected by the intrinsic fine-scale source structure:
\begin{equation} 
    \theta_{\rm{cpt}} \gtrsim \sqrt{\frac{4\ln2 \ln \left( \frac{|\mathcal{V}_{\rm{SMT-LMT}}|}{|\mathcal{V}_{\rm{ALMA-LMT}}|}\right)}{\pi^2(|\bf{u}|_{\rm{ALMA-LMT}}^2-|\bf{
u}|_{\rm{SMT-LMT}}^2)}}.
\end{equation}
Here, $\mathcal{V}_{i-j}$ 
denotes the true visibility on the baseline $i-j$. The ratio of the true visibility amplitudes is lower-bounded by 
\begin{equation}
    \frac{|\mathcal{V}_{\rm{SMT-LMT}}|}{|\mathcal{V}_{\rm{ALMA-LMT}}|} \geq  \left(1-\sqrt{\Delta g_{\rm{ALMA}}^2+\Delta g_{\rm{SMT}}^2}\right)  \frac{|V_{\rm{SMT-LMT}}|}{|V_{\rm{ALMA-LMT}}|},
\end{equation}
where $V_{i-j}$ and $\Delta g_i$ denote the calibrated measured visibility and the deviation of its residual station gain from unity, respectively, after gain calibration using calibrators.
$\Delta g_{\rm{ALMA}}$ and $\Delta g_{\rm{SMT}}$ are estimated to be 0.05 and 0.06 respectively, corresponding to the standard deviation of gain solutions from calibrator data across multiple imaging methods described in \S\ref{sec:gainerrors}.  
We take the median of the visibility amplitude ratio from the collection of constraints derived for each single VLBI scan to derive a robust estimate of the minimum source size.
With this method, we are able to derive a conservative constraint, which is agnostic to the assumed source structure (e.g., the presence of a null from  a ring-like structure).


{\it Constraint 2.} 
The second constraint comes from the curvature of visibility amplitudes between the intra-site baselines and the LMT--SMT baseline. Since the compact flux density should not exceed the total flux density measured with the intra-site baselines, the amplitude fall from the intra-site to LMT--SMT baselines gives the maximum limit of that from the compact flux density to LMT--SMT baseline. Therefore, it gives the maximum limit of the source FWHM size with an equivalent circular Gaussian as
\begin{equation}
    \theta_{\rm{cpt}} \leq \frac{2 \sqrt{\ln2}}{\pi |\bf{
    u}|} \sqrt{\ln \frac{F_{\rm{tot}}}{|\mathcal{V}_{\rm{LMT-SMT}}|}}.
\end{equation}
With the residual gain uncertainty, this maximum limit is upper-bounded by
\begin{equation}
    \frac{1}{|\mathcal{V}_{\rm{LMT-SMT}}|} \leq \frac{1+\sqrt{\Delta g_{\rm{LMT}}^2+\Delta g_{\rm{SMT}}^2+\Delta g_{\rm{tot}}^2}}{|V_{\rm{LMT-SMT}}|}.
\end{equation}
In the same manner as with other stations, $\Delta g_{\rm{LMT}}$ is estimated to be 0.12 post calibrator gain-transfer. $\Delta g_{\rm{tot}}$, the fractional uncertainty in the light curve, is estimated to be at most 0.2 based on the consistency of light curves obtained with different instruments / reduction pipelines \citep[see][]{Wielgus2022}. 
Similarly to Constraint 1, the median of the ratio is adopted to mitigate the effects of statistical errors.

{\it Constraint 3.}
The minimum compact flux density can be derived by the maximum amplitudes of LMT--SMT baseline, since the visibility amplitudes are maximum at zero baseline length, as given by
\begin{align}
    F_{\rm{cpt}} &\geq |\mathcal{V}_{\mathrm{SMT-LMT}}|\\
    &\geq (1-\sqrt{\Delta g_{\rm{LMT}}^2+\Delta g_{\rm{SMT}}^2})|V_{\rm{LMT-SMT}}|.
\end{align}
In the above equation, the equality is satisfied in the extreme situation when the source is completely unresolved at the LMT--SMT baseline, whereas in reality it partially resolves the source structure. A stronger constraint is given by the equivalent circular Gaussian with the minimum source size (Constraint 1) extrapolated from the LMT--SMT baselines, described by,
\begin{equation}
    F_{\rm{cpt}} \geq |\mathcal{V}_{\rm{LMT-SMT}}| e^{\frac{(\pi |\bf{u}| \theta_{\rm{cpt}})^2}{4 \ln 2 }}.
\end{equation}

We note that the upper and lower limits on the source size obtained from Constraints 1 and 2 are along the directions of LMT-SMT and LMT-ALMA baselines. Using synthetic data from general relativistic magnetohydrodynamics (GRMHD) and semi-analytic geometric models in \citetalias{PaperIII}, we have verified that these size constraints remain valid for various morphologies sharing similar profiles of visibility amplitudes with EHT \sgra data (see \citetalias{PaperIII} for details).

The derived constraints are summarized in \autoref{tab:source_size} and \autoref{fig:source_size} (a) for each observing day and for both days combined. As shown in \autoref{fig:source_size} (b), the equivalent Gaussians with the minimum and maximum sizes reasonably give the upper and lower bounds of the visibility amplitudes, respectively, on short baselines. 
Either of the source size constraints indicate that the 2017 EHT \sgra data resolve compact emission of several tens of microarcseconds, consistent with the early EHT observations \citep{Doeleman08, Fish_2011, Johnson_2015, Lu_2018, 2018ApJ...865..104J}.

The estimated minimum fractional compact flux density is $\sim 70-90$\,\%, indicating that the vast majority of the radio emission from \sgra arises from the horizon-scale emission resolved with the EHT. 
The maximum limit on the extended emission resolved out on the shortest VLBI baselines is indeed much less than for \m87, 3C\,279, and Centaurus~A---all of which are known to have a prominent extended jet (\citetalias{M87PaperIV}; \citealt{Kim_2020}; \citealt{2021Janssen})---as expected by the non-detection of an extended jet in longer-wavelength VLBI observations across many decades (see \S\ref{sec:scattering} 
and references therein) and early EHT observations. 

\subsubsection{Source variability}


Studies of the variability of \sgra light curves at 230\,GHz in 2005--2017 \citep[e.g.,][]{Marrone08,Dexter14,Bower2018} indicate that on long timescales (from a few days to years) the source fluctuates between 2.0 and 4.5\,Jy. During the 2017 EHT observations \sgra was in a low luminosity state with a mean flux density of $\sim2.4\pm 0.2$\,Jy between $213-229$ GHz (\autoref{tab:sed_2017}). The two days for which VLBI data analysis is presented in this series of papers exhibit a low degree of variability that is typical for this source, characterized by the modulation index (standard deviation divided by mean) $\sigma/\mu <10$\%. The power spectral density has a red noise character on timescales from $1$\,min to several hours, with a transition to white noise for longer timescales.
On 2017 April 11 the millimeter flux rises by over 50\%, following the X-ray flare maximum with a delay of $\sim$2\,h. The data on this day appear significantly more variable. An analysis of the 2017 April 11 VLBI data will be presented in a separate future publication. A detailed analysis of the full data set of \sgra light curves contemporaneous with the EHT 2017 observations is presented in the companion paper \citet{Wielgus2022}.

\begin{figure}
\includegraphics[width=\columnwidth]{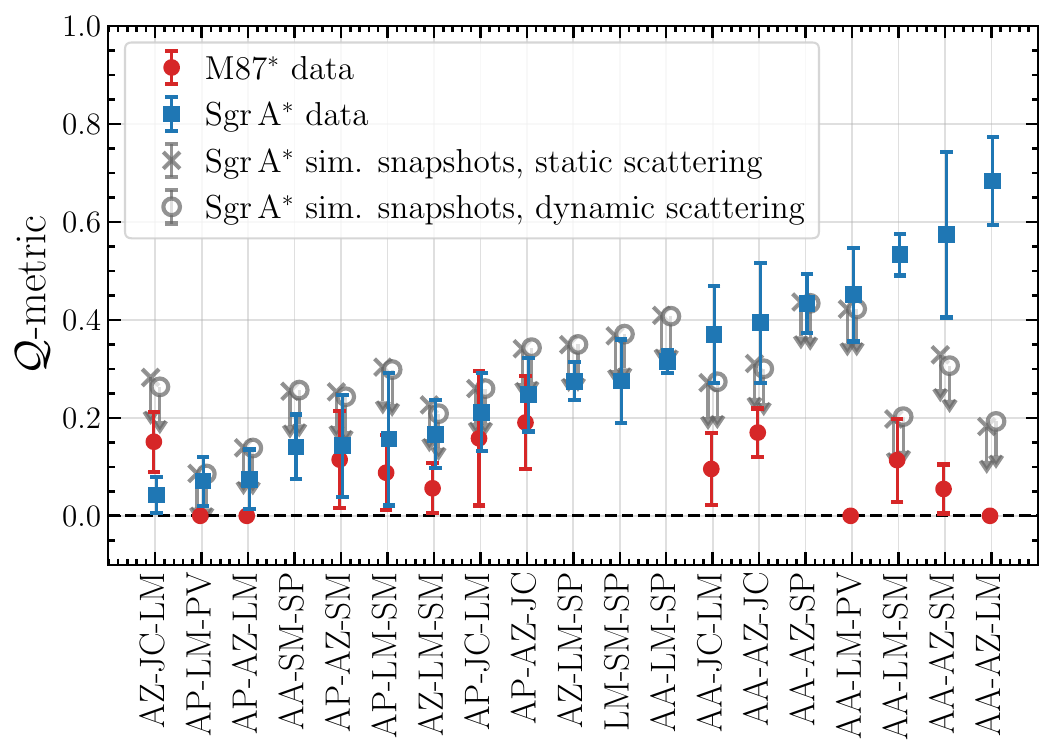}
\caption{$\mathcal{Q}$-metric values (\autoref{eq:qmetric}) indicating intrinsic closure phase variability on baseline triangles with mutual visibility of both \m87 and \sgra in ascending order of the \sgra variability. The two-letter station codes are taken from \autoref{tab:ER6uncertainties}. Only closure phase tracks with more than ten data points after averaging down to 120\,s were used. The closure phase tracks were detrended within two hour segments by subtracting a third-degree polynomial before computing the $\mathcal{Q}$-metric. The plotted values are averages across days (2017 April 5, 6, 10, and 11 for \m87, 2017 April 6 and 7 for \sgra) and bands, with the error bars indicating the 1$\sigma$ spread across the days and bands. The $\mathcal{Q}$-metric for the \m87 and \sgra data are shown in red and blue, respectively. The grey points show 1$\sigma$ upper limits of the distributions of $\mathcal{Q}$-metric values obtained from synthetic data generated from intrinsically static source models, with either a static or moving scattering screen with parameters from \citet{2018ApJ...865..104J}, but with an enlarged power-law index of $\alpha=1.5$ for the interstellar turbulence to obtain more conservative upper limits of scattering-induced variability. The static source models are a mix of models from the GRMHD library, which contains a large range of source structures, and the models used to calibrate the imaging methods in \citetalias{PaperIII}. Upper limits are computed based on the maximum $\mathcal{Q}$-metric values from independent realizations of possible source models and thermal noise on each triangle.
}
\label{fig:qmetric}
\end{figure}

The $\mathcal{Q}$-metric \citep{Roelofs2017} quantifies the variability of a compact source resolved on VLBI baselines in a series of closure phases as a function of time on a specific baseline triangle. The metric compares the observed closure phase variations ($\hat{\sigma}^2$) to those expected from thermal noise ($\tilde{\epsilon}^2$), after detrending the data to account for slow variations due to the evolution of baselines as the Earth rotates \citep[see][for details]{Roelofs2017}:
\begin{equation}
    \mathcal{Q} = \frac{\hat{\sigma}^2 - \tilde{\epsilon}^2}{\hat{\sigma}^2}\,.
\label{eq:qmetric}
\end{equation}

A static source is expected to give a $\mathcal{Q}$-metric value close to zero, and a variable source can give a $\mathcal{Q}$-metric value up to one.

\autoref{fig:qmetric} shows the $\mathcal{Q}$-metric for \m87 and \sgra data after dividing the closure phases into two-hour segments and detrending those with a third-order polynomial. The $\mathcal{Q}$-metric is generally higher for \sgra than for \m87, which does not necessarily indicate that \sgra varies more strongly because of the different error budgets for the two sources. However, the different $\mathcal{Q}$-metric values do indicate that intrinsic variability is detected with higher significance for \sgra than for \m87. The measured variability occurs on timescales between the visbility averaging time of 120\,s and the two hour long detrending segments. The high sensitivity to thermal noise can lead to different $\mathcal{Q}$-values on redundant triangles.  

Because of imperfections in the detrending procedure (a third-order polynomial does not generally capture any static source structure) and the particular realization of the thermal noise, a static source may also give nonzero $\mathcal{Q}$-metric values. In order to determine on which triangles we detect significant variability, the measured $\mathcal{Q}$-metric values are therefore compared to those simulated for a range of static source models for \sgra. Shown in \autoref{fig:qmetric} are 1$\sigma$ upper limits of the distributions of $\mathcal{Q}$-metric values obtained from synthetic data generated from different single static GRMHD-GRRT snapshots \citepalias{PaperV}, with different realizations of either a static or moving scattering screen. We have applied random position angle rotations to 12 random frames drawn from MAD/SANE models with spins of $\pm 0.94, \pm 0.5, 0$, $R_\mathrm{high}= 10, 40$ and inclination angles of $\ang{10}, \ang{30}, \ang{50}, \ang{70}$ \citepalias[a description of these models and their parameters is given in][]{PaperV}. For these 960 different source structures, we have generated synthetic data using 100 different thermal noise realizations for each observing day and the low and high frequency bands. We applied the same procedure to eleven frames of each of the source models used to calibrate the imaging methods in \citetalias{PaperIII}, which were chosen to match the measured visibility amplitudes of \sgra. The resulting $\mathcal{Q}$-metric values were bootstrapped to obtain the same number of values as those from the GRMHD models.

Scattering-screen variability generally has a small effect on the $\mathcal{Q}$-metric. Comparing the measured and simulated $\mathcal{Q}$-metric values, we see a significant excess for \sgra on ALMA-LMT-SMA, ALMA-SMT-SMA, and ALMA-SMT-LMT. For these triangles, the observed closure phase variability cannot be explained by variability due to interstellar scattering or imperfections in the detrending procedure. The excess variability in degrees $\breve{\mathcal{Q}}=\sqrt{\hat{\sigma}^2 - \tilde{\epsilon}^2}$ is \ang{16.4}, \ang{6.8}, and \ang{3.1} for these triangles, respectively. Hence, although the $\mathcal{Q}$-metric does indicate intraday closure phase variability in the \sgra data, it only does so on a few triangles, and the variability amplitude is small.

More detailed discussions of the source variability based on static and time-dependent reconstructions of the underlying \sgra source model, as well as examples of typical closure phase trends, are given in \citetalias{PaperIII, PaperIV}.

\subsection{Multi-wavelength Data Products}
\label{subs:sup_data}

\subsubsection{mm, NIR, and X-ray data products}
\label{subs:mwl_data}

The supplementary ground- and space-based data products leveraged here are published in original works and/or are available in public archives. 

EAVN observations of \sgra\ at 22 and 43 GHz are available in \citet{Cho_2022}, GMVA 86 GHz observations are published in \citet{2019ApJ...871...30I}, and a detailed discussion of the VLT/GRAVITY NIR flux distribution can be found in \citet{GRAVITY2020}.

\chandra, \swift, and \nustar\ data products are available via NASA archives\footnote{The \chandra Data Archive: \url{https://cxc.harvard.edu/cda/}.}\textsuperscript{,}\footnote{The \swift Data Archive: \url{https://swift.gsfc.nasa.gov/archive/}.}\textsuperscript{,}\footnote{The \nustar Data Archive: \url{https://heasarc.gsfc.nasa.gov/docs/nustar/nustar_archive.html}.} and are collected in  
the \ac{eht} data portal\footnote{\url{https://eventhorizontelescope.org/for-astronomers/data}} under the \texttt{2021-DXX-XX} code.
The repository contains the following data products: 
\begin{enumerate}
\item Description of observations and data processing (format: text).
\item Fluxes from Swift-XRT observations (format: CSV).
\item Fluxes from Chandra observations (format: CSV).
\item Fluxes from NuSTAR observations (format: CSV).
\item Scripts, spectral, and response files for modeling Chandra and NuSTAR data (format: standard X-ray data formats).
\item Sampled posterior distributions of X-ray spectral model based on Chandra and NuSTAR data (format: FITS).
\end{enumerate}

The EHT data portal also contains the broadband spectrum table (see \S\ref{subs:sgra_sed} below) with frequency, flux density, its uncertainty, and instrument index (format: CSV).






\subsubsection{Time-averaged spectral energy distribution}
\label{subs:sgra_sed}

\begin{figure*}
\includegraphics[width=\textwidth]{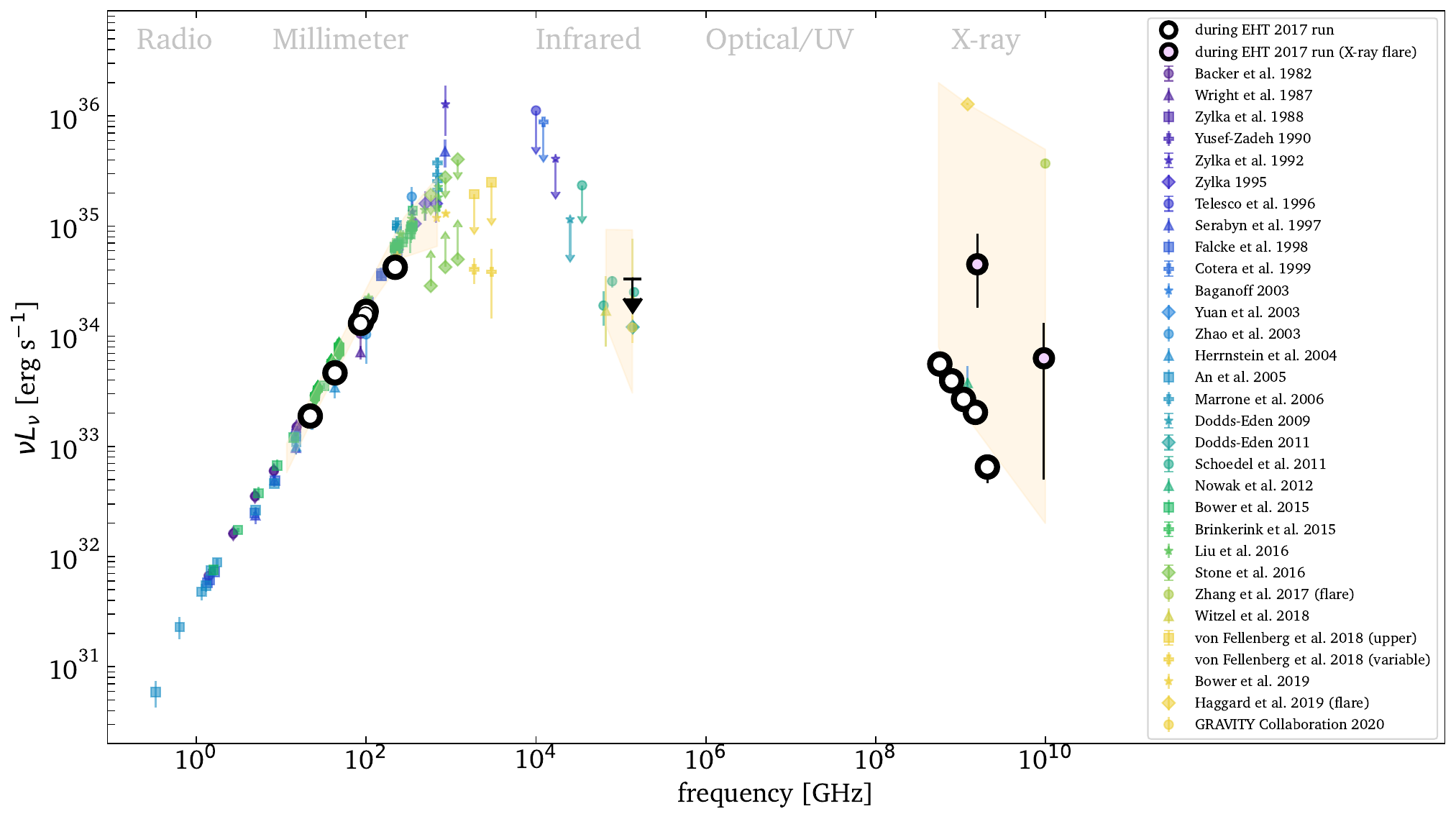}
\caption{
The time-averaged SED for the compact \sgra source during the 2017 EHT run is shown as black open circles and a NIR upper limit. Luminosities for the X-ray flare observed on 2017 April 11 are indicated as black circles filled with light purple. Table \ref{tab:sed_2017} lists the 2017 values in units of flux and luminosity. Colored background points display the historic SED of \sgra in flaring and quiescent states with the light yellow polygons indicating the range of previously observed variability \citep{Backer1982,Zylka1988,Wright1987,Yusef-Zadeh1990,Zylka1992,Zylka1995,Telesco1996,Serabyn1997,Falcke1998,Cotera1999,Baganoff2003,Yuan03,Zhao2003,Herrnstein2004,An2005,Marrone2006,Dodds-Eden2009,Dodds-Eden2011,Schodel2011,Nowak12,Bower2015,Brinkerink2015,Liu2016,Stone2016,Zhang2017,Witzel2018,Haggard2019,vonFellenberg2018,Bower2019,GRAVITY2020}. A time-binned historic SED representative of states without extreme variability is presented in \citetalias{PaperVI}.
\label{fig:sed_2017}}
\end{figure*}

\autoref{fig:sed_2017} displays the SED for \ac{sgra} during the 2017 EHT campaign (open black circles) over-plotted on the historically observed broadband spectrum (colored points). The SED illustrates \ac{sgra}'s wide range of variable and non-variable states. 
Larger shaded swaths mark regions of the SED where the source is particularly variable; their bounds mark characteristic quiescent emission and high flux states, which can last for timescales of minutes to hours. (A time-binned historic SED representative of states without extreme variability is presented in \citetalias{PaperVI}.)
We do not plot a 
typical quiescent value for \nustar frequencies (3--79\,keV) since upper-limits are complicated by contributions from non-\sgra sources in the galactic center.
The flux and luminosity values from each observatory coordinating during the EHT campaign are listed in Table \ref{tab:sed_2017}. 
The quiescent and flare X-ray luminosities in the SED are $\nu$L$_{\nu}$ in units of erg s$^{-1}$; their equivalent $\nu$F$_{\nu}$ values in erg s$^{-1}$cm$^{-2}$ (Table \ref{tab:sed_2017}) differ slightly from the integrated flux values in Table \ref{tab:xrayflares}. 

During the EHT run, \sgra's SED is consistent with historical observations of the black hole in the radio, mm, NIR, and X-ray, as outlined in the Introduction and \S\ref{sec:sgra-mwl-var}. For example, the moderately bright X-ray flare detected with \chandra and \nustar on 2017 April 11 falls within the range of previously observed X-ray flares (see \S\ref{subs:space_obs} and \autoref{fig:sed_2017}). 
Since \sgra is not in an exceptional state during the 2017 EHT campaign, these observational constraints 
on the broadband SED 
offer valuable priors on theoretical models aiming to constraint GR, for example the Kerr metric investigation presented in \citetalias{PaperVI}.

\citetalias{PaperV} in this series also uses these (quasi-)simultaneous EHT 1.3\,mm and multi-wavelength constraints (on both luminosity and degree of variability) to aid in model selection and to provide a physical interpretation for these data, for example by testing aligned, tilted, and stellar wind-fed scenarios using time-dependent GRMHD models. They compare specifically to three MWL bands, 86 GHz, 2.2\,$\mu$as, and X-ray, which are relatively independent and thus probe different physics, and find that no one existing model can meet all of the EHT and MWL constraints. 
These challenges motivate improvements to the GRMHD model suites, and encourage additional, joint EHT and MWL campaigns to fully characterize \sgra's short and long term variability. 

\subsubsection{Characteristic multi-wavelength variability}
\label{sec:sgra-mwl-var}

\sgra's multi-wavelength variability has been studied in detail for more than two decades. 
Substantial obscuration in the plane of the Galaxy blocks our view of its optical and UV emission, but observations at other wavelengths show flickering, flares, and other stochastic processes on long and short timescales \citep[e.g.,][]{Baganoff01, Dodds-Eden09, Nowak12, Witzel12, Neilsen15, Bower2015, Haggard2019, Do19b, GRAVITY2020}.
Theoretical models also anticipate that observations at different wavelengths 
will probe different size scales and resolutions \citep[e.g.,][]{FalckeMarkoff2013}.
We briefly describe \sgra's characteristic MWL variability, to put the X-ray flares (\S \ref{sec:joint-xray}) and the broadband spectrum (\S \ref{subs:sgra_sed}) observed in 2017 into this broader context. 

\sgra's X-ray flux distribution can be decomposed into a steady quiescent component described by a Poisson process,  
and a variable power law component attributed to  non-thermal flares that appear approximately once per day \citep{Neilsen15}. 
\sgra's quiescent NIR light curves show a red noise process and a non-linear, non-Gaussian flux density distribution, skewed to higher flux densities, which  
changes slope 
near a median flux density of $1.1\pm0.3$ mJy \cite[flux densities below 0.1 mJy are rarely observed; e.g.,][]{Witzel12,Do19b,GRAVITY2020}. Thus the NIR flux can also be attributed to separate quiescent and variable components.

The X-ray and NIR variability timescales are both typically several hours, consistent with \ac{sgra}'s ISCO period of 
approximately 4--30\,min for prograde orbits (vs.\ 5 days to 1 month for \m87; for further discussion of this comparison, see \citetalias{PaperI}).
Recently, the orbital motion of a compact polarized hot spot just outside of the ISCO has been offered as an explanation for high resolution, time resolved interferometric NIR observations \citep{2018Gravity}, reinforcing the notion that this may be a horizon-scale phenomena. 

Correlations between the X-ray and NIR flux peaks \citep[e.g.,][]{Dodds-Eden09, Ponti17, Boyce19, Gravity2021, Michail2021, Boyce2022} and similarities in their statistical properties and flux density distributions
\citep[][]{Witzel12,Witzel18,Witzel2021, Neilsen15} point to a physical connection between the emission at these wavelengths, 
%
%
though the X-ray structure function seems to have less power at short timescales than the NIR structure function \citep{Witzel12,Witzel2018,Neilsen15,Witzel2021}. 

At submillimeter and radio wavelengths \ac{sgra} also shows a quiescent state superposed with almost continuous variability \citep[e.g.,][]{Miyazaki2004,Macquart06,YusefZadeh2011,Brinkerink2015}. 
These longer-duration flares may be delayed by a few hours relative to the X-ray/NIR flares, but simultaneous observations are sparse and the correlations remain tenuous \citep{Capellupo17}. 
Some submm flares have also been associated with NIR flares, while others show no correlation \citep{Marrone08,Morris12,Fazio18,Michail2021, Boyce2022}. 
\citet{Iwata2020} pursued a detailed study of \ac{sgra}'s flux density distribution at 217.5, 219.5, and 234.0\,GHz, finding variability on timescales of $\sim$10's of minutes to hours that is likely to arise near the ISCO. 
They find no lag between 217.5 and 234.0 GHz. 

Analysis by \citet{Wielgus2022} based on the ALMA and SMA data associated with this 2017 EHT campaign show \sgra mostly in a quiescent state at 213, 220, and 229\,GHz (Figs. \ref{fig:mwl_cover} and \ref{fig:sed_2017}), though the 2017 April 11 observations following the X-ray flare show enhanced millimeter wavelength variability. The mm light curves are consistent with a red noise process with the power spectral density slope between $-2$ and $-3$ on timescales between 1 minute and several hours.  
%
In an independent study, \citet{Bower2018} detect both linear and circular polarization in 
\ac{sgra}'s millimeter emission. They find a mean rotation measure (RM) of $\sim -5\times10^5\,{\rm rad} \,{\rm m}^{-2}$ and variability on timescales of hours to months. 
Long-term variability in the RM (of order weeks to months) is likely due to turbulence in the accretion flow, while short-term variability seems to arise from complex emission and propagation effects near the black hole. They also detect circular polarization with a mean value of $-1.1\pm0.2$\% that is variable on timescales of hours to months \citep{Bower2018}.
EHT polarimetric measurements of \sgra will be the subject of future works in this series. 

These and other MWL observations have led to broad consensus that near \sgra's event horizon there are two components: (1) a relatively stable quiescent accretion flow, and (2) flares or bright flux excursions that vary on shorter timescales. 
Accretion models for \sgra suggest that the system may drive a jet, and some data support this possibility. For example, an observed increase of variability amplitude with frequency and persistent time lags in the $18\,$--$\,43$\,GHz band may be indicative of a jet outflow \citep{2021Brinkerink}.
Yet despite this and other tentative claims at $\gamma$-ray, X-ray, and radio wavelengths \citep[e.g.,][]{SuFinkbeiner2013,Li2013,Li2019,YusefZadeh2012,Rauch2016}, a jet has not yet been conclusively confirmed.  
The accretion flow and the jet base are both candidates to drive \sgra's variability (e.g., the base of the jet very likely dominates the emission near \m87), yet while their timescales may be distinct, 
the base of a jet and a variable RIAF remain difficult to tell apart (see \citetalias{PaperV}).


Interstellar scattering may induce additional variable signals over timescales similar to (or longer than) those from the quiescent accretion flow. For example, one explanation for the detectability of the GC magnetar was a variable scattering screen \citep{Eatough2013,Bower2014}. 
Meanwhile, interactions with orbiting objects may cause additional variability, e.g., disruptions of $\sim$km-sized rocky bodies, gaseous structures, or other interlopers \citep{Cadevz2008, Kostic2009, Zubovas2012, Gillessen12, Ciurlo2020, Peissker2021}. Contemporaneous observations with EHT and MWL facilities, particularly simultaneous capture of serendipitous events like flares, continue to offer the best opportunity for disentangling these contributions to \ac{sgra}'s variable emission. 

\section{Summary and Conclusions}
\label{sec:concl}

\sgra and \m87 are the primary targets of the \ac{eht}, for which the April 2017 observing campaign provided the required resolution and sensitivity to obtain horizon-scale images of these supermassive black holes.
\m87 total intensity and polarization results from the 2017 data have been published earlier \citepalias{M87PaperI, M87PaperII, M87PaperIII, M87PaperIV, M87PaperV, M87PaperVI, M87PaperVII, M87PaperVIII}. 
In this new series of papers \citepalias{PaperI, PaperII, PaperIII, PaperIV, PaperV, PaperVI}, we present the imaging, analysis, and interpretation of the 2017 \sgra \ac{eht} and accompanying data at X-ray, NIR, and millimeter wavelengths from EAVN, GMVA, VLT/NACO, \swift, \chandra, and \nustar.
%
%
\sgra was observed by the \ac{eht} on 2017 April 5, 6, 7, 10, and 11. In this series of papers, we focus our analysis on the days with the best $(u, v)$-coverage, 2017 April 6 and 7; a detailed analysis of the full set of \ac{sgra} millimeter light curves can be found in the companion paper by \citet{Wielgus2022}. 

We use the two independent pipelines EHT-HOPS \citep{2019ApJ...882...23B} and rPICARD \citep{2018evn..confE..80J, 2019A&A...626A..75J} for the calibration of the \ac{eht} data.
By utilizing the data from both pipelines, we obtain scientific results in a robust manner, independent of assumptions made for the data calibration.
With various jackknife tests, we estimate the amount of systematic noise present in the EHT data and verify that data issues are mitigated.

We present updates to the a priori gain calibration for several \ac{eht} stations.
Additionally, we perform a network calibration \citep{2019ApJ...882...23B} based on the total flux of the source to correct the gains of the co-located \ac{eht} stations.
To account for the rapid flux variability of \sgra in this process, we use light curves obtained at a short intra-scan cadence from the \ac{aa} and \ac{sm} connected element interferometers.
Finally, we transfer gain solutions from self-calibration solutions of the co-observed J1924$-$2914 and NRAO\,530 calibrator sources.
With the time-dependent network calibration and calibrator gains transfer, it is possible to partially correct for time-variable station gain errors that are difficult to solve for using the \sgra data alone due to intrinsic source variability.

The calibrated visibility amplitudes can be described by a blurred ring with a diameter of \sdiam 
(a deeper discussion of the source size and morphology can be found in \citetalias{PaperIII} and \citetalias{PaperIV}). 
The majority of the total flux measured in our \ac{vlbi} experiment arises from horizon scales in \sgra. Using multi-wavelength constraints on the interstellar scattering screen towards \sgra, we show that the scattering-induced angular broadening of the source is sub-dominant to the intrinsic source structure uncovered by the \ac{eht}.
The refractive noise added by the scattering screen is only relevant for data at baseline lengths $\gtrsim 6$\,G$\lambda$.

Without making strong assumptions about the source structure, we find a compact source size of 39\,--87\,\ac{muas} for \sgra, which is in agreement with earlier 1.3\,mm \ac{vlbi} observations of the source. A more precise modeling-based estimation of the source size is given in \citetalias{PaperIII}.
On 2017 April 6 and 7, the source was in a low luminosity state, where the total flux fluctuates around 2.4\,Jy with a modulation index of less than $10\,\%$.
The \sgra closure phases show clear intrinsic structural variability on timescales of a few minutes to a few hours that is further investigated in \citetalias{PaperIII, PaperIV}.

Multi-wavelength observations show that \sgra was in a mostly quiescent state, with broadband flux levels consistent with historical measures. We detect two X-ray flares: one very faint flare on 2017 April 7 and a brighter flare on 2017 April 11. The fainter 2017 April 7 flare is detected at low significance by \swift\ and \chandra, and the brighter 2017 April 11 flare is detected more confidently by \chandra\ and \nustar. 
ALMA and SMA 2017 April 11 observations begin immediately after the bright X-ray flare and show enhanced millimeter wavelength variability \citep{Wielgus2022}. 
These multi-wavelength data offer important constraints for theoretical models; indeed, no one of the GRMHD models presented in \citetalias{PaperV} can match the full suite of multi-wavelength constraints. These unprecedented EHT and MWL data thus provide a rich opportunity to improve models of \ac{sgra} and to advance our understanding of the physics near the SMBH event horizon. The high quality, nearly simultaneous multi-wavelength SED is additionally valuable for understanding the priors for sophisticated tests of GR (see \citetalias{PaperVI}). Looking ahead, future detailed analysis of the EHT and MWL observations on 2017 April 11 holds great promise for understanding the underlying mechanisms that drive \sgra's flares and other variability.



\acknowledgments


\facilities{EHT, ALMA, APEX, IRAM:30m, JCMT, LMT, SMA, ARO:SMT, SPT, Chandra, EAVN, GMVA, NuSTAR, Swift, VLT.}
\software{DiFX \citep{2011PASP..123..275D}, CALC, PolConvert \citep{martividal2016}, HOPS \citep{2004RaSc...39.1007W}, EHT-HOPS Pipeline \citep{2019ApJ...882...23B}, CASA \citep{2007ASPC..376..127M}, rPICARD \citep{2018evn..confE..80J, 2019A&A...626A..75J}}

\bibliography{main.bib, EHTCPapers}

\end{document}